\documentclass[twocolumn,floats,floatfix,aps,pra]{revtex4}

\usepackage{amsfonts}
\usepackage{amssymb}
\usepackage{amsmath}
\usepackage{calc}
\usepackage[dvips]{graphicx}
\usepackage{bm}
\usepackage{mathrsfs}
\usepackage{float}

\def\be{\begin{equation}}
\def\ee{\end{equation}}
\def\bea{\begin{eqnarray}}
\def\eea{\end{eqnarray}}
\def\bse{\begin{subequations}}
\def\ese{\end{subequations}}

\def\1{\mathbf{1}}

\begin{document}

\author{Lachezar S. Simeonov}
\affiliation{Department of Physics, Sofia University, James Bourchier 5 blvd, 1164 Sofia, Bulgaria}

\title{Derivation of Schr\"{o}dinger's equation from the Hamilton-Jacobi and the Eikonal equations}
\date{\today }

\begin{abstract}
Most authors of textbooks on quantum mechanics either postulate or sketch a short `ad hoc` derivation of Schr\"{o}dinger's equation. In this work we give a detailed derivation of Schro\"{o}dinger's equation from the Hamilton-Jacobi equation and the Eikonal equation in geometrical optics. We start from the historical debates on the nature of light - whether it is a beam of particles, or waves in the aether. We derive the Eikonal equation and show the conditions when a wave can behave as a beam of particles. Then we discuss several experiments with an electron gun, that show clearly diffraction and interference of a single electron. Next, in order to explain these experiments, we derive Schr\"{o}dinger's equation by comparing Hamilton-Jacobi equation in classical mechanics with the Eikonal equation in geometrical optics. To do that, we first show how to derive the wave equation from the Eikonal equation (not the other way around!). Second, we use this method to derive Schr\"{o}dinger's equation from the Hamilton-Jacobi equation. Next, we \textit{derive} Born's statistical rule using the early understanding of de Broglie that \textit{both} particles and waves exist. Afterwards, we show that historically people preferred to remove the particles (as well as their trajectories) altogether from de Broglie's ideas but retained Born's rule (the so called Copenhagen interpretation). These derivations of the foundations of quantum mechanics do \textit{not} follow \textit{precisely} the history of the subject. Rather we select some early ideas and experiments in a judicious manner to present Schr\"{o}dinger's equation in a logical and ordered way. We use the electron gun experiments instead of black body radiation and photoelectric effect. Our derivation may bring more light and satisfaction for the undergraduate students about the confusing and rather mysterious subject of quantum mechanics.
\end{abstract}


\maketitle

\section{Introduction}
Many textbooks on quantum mechanics do \textit{not} derive Schr\"{o}dinger's equation but simply postulate it (for instance \cite{Griffiths2018} and \cite{Tannoudji2019}) and proceed directly to applications. Other textbooks use de Broglie's idea that free particles with momentum $\textbf{p}$ and energy $E$ are `associated` with plane waves: $\Psi(\textbf{r},t)= Ae^{i(\textbf{p}\cdot \textbf{r}-Et)/\hbar}$, where $\hbar$ is Planck's constant and $A$ is the amplitude of the plane wave. Next, they state Born's rule. Then they use the correspondence principle as well as the principle of superposition, and by building wave packets, they \cite{Bohm1989, Messiah2014, Greiner2001, Dirac1959} reach the conclusion that to obtain the average momentum, we need to calculate the integral:
\begin{equation}
\left\langle\textbf{p}\right\rangle=\int_{V_{\infty}}\Psi^{*}(\textbf{r},t)\left(-i\hbar\nabla\right)\Psi(\textbf{r},t) dV
\end{equation}
where $V_{\infty}$ is the whole of space. This `derivation` of the momentum operator $\hat{\textbf{p}}=-i\hbar\nabla$ is similar to what is done in \cite{Landau1981}, where the authors derive the momentum operator from symmetry considerations. Once the operator of momentum $\hat{\textbf{p}}$ is derived, some authors \cite{Griffiths2018, Greiner2001} simply postulate that for a general power $n$:
\begin{equation}
\left\langle\textbf{p}^{n}\right\rangle =\int_{V_{\infty}}\Psi^{*}(\textbf{r},t)\hat{\textbf{p}}^{n}\Psi(\textbf{r},t) dV
\end{equation}
Then, they `derive` the Hamiltonian: $\hat{H}=\hat{\textbf{p}}^{2}/2m+U=-\hbar^{2}\nabla^{2}/2m+U$ and postulate Schr\"{o}dinger's equation
\begin{equation}
i\hbar\frac{\partial\Psi}{\partial t}=\hat{H}\Psi
\end{equation}
Other authors \cite{Bohm1989, Messiah2014} use the correspondence principle and guess what the Schr\"{o}dinger's equation ought to be. Others \cite{Landau1981, Dirac1959} reach the conclusion that in the semi-classical regime, the wave function $\Psi$ is:
\begin{equation}
\Psi=Ae^{iS/\hbar}
\end{equation}
where $S$ is the action in classical mechanics. After some semi-classical arguments \cite{Landau1981} and \cite{Dirac1959} guess Schr\"{o}dinger's equation.

In this paper we derive in detail Schr\"{o}dinger's equation by comparing Hamilton-Jacobi equation and the Eikonal equation in geometrical optics \cite{Sommerfeld1964optics, Born1999}. We first show the conditions when a wave behave as a beam of particles. These conditions follow from our derivation of the Eikonal equation from the wave equation. Then we argue that Hamilton-Jacobi equation \textit{is} a type of Eikonal equation, which should follow from some unknown wave equation. Next, we invent a specific mathematical procedure that allows us to derive the wave equation from the Eikonal equation (not the other way around!). We use this procedure to derive Schr\"{o}dinger's equation from Hamilton-Jacobi equation. As far as the author knows, this particular derivation has not been made before. People have only speculated on the similarity between the Eikonal equation and the Hamilton-Jacobi equation \cite{Landau1981,Dirac1959} but to our knowledge, a specific derivation of this type has not been made. 

This paper is organized as follows. In Section II we give a very brief history of the corpuscular vs wave theory of light. In Section III we show when waves behave as a beam of particles. In Section IV we derive the time-dependent Eikonal equation. In Section V we consider three electron gun experiments, which clearly show diffraction and interference of single electrons. In Section VI we derive Hamilton-Jacobi equation in classical mechanics. In Section VII we compare Hamilton-Jacobi equation and the Eikonal equation (Table I) and we invent a procedure which allows us to derive the wave equation from the Eikonal equation. We use this procedure to derive Schr\"{o}dinger's equation from the Hamilton-Jacobi equation. In Section VIII we continue with the history of the Schr\"{o}dinger's equation and compare the original de Broglie's idea (pilot wave theory) with the Copenhagen interpretation. We show that people discarded the particles and their trajectory, but postulated collapse of the wave function. For completeness sake we derive Born's rule in pilot wave theory and show that pilot wave theory can also explain the three electron gun experiments. Section IX gives the conclusions.

\section{What is light - \textit{very} short history}

\subsection{Light as particles}
One of the most influential persons in the history of physics is Descartes. Perhaps his major contribution was his idea that all of physics can be reduced to mechanics \cite{Whittaker1989}, including optics and gravity. For instance, his understanding of the solar system was that the empty space was filled with an invisible substance called `aether`. The sun was considered to be at the center of a giant aether vortex, which sweeps along the planets. Descartes and his followers had employed this idea to understand gravity, electricity, magnetism and optics.

When it comes to light, people devised two competing theories - light was considered either as particles (corpuscules) or as vibrations in the aether, similar to the sound, which is vibrations in the air. 

Newton (though he was a believer in the aether \cite{Whittaker1989}) considered light as particles, not waves. His reasoning was that it is very difficult to explain the rectilinear motion of light, if it was waves in the aether. Indeed, let us consider a light beam that travels through a small aperture as shown in FIG. 1. Newton said that light never entered the geometrical shadow (he was wrong about that). This rectilinear motion of light resembles a beam of particles, not waves. 
\begin{figure}[tb]
\includegraphics[width= 1.0\columnwidth]{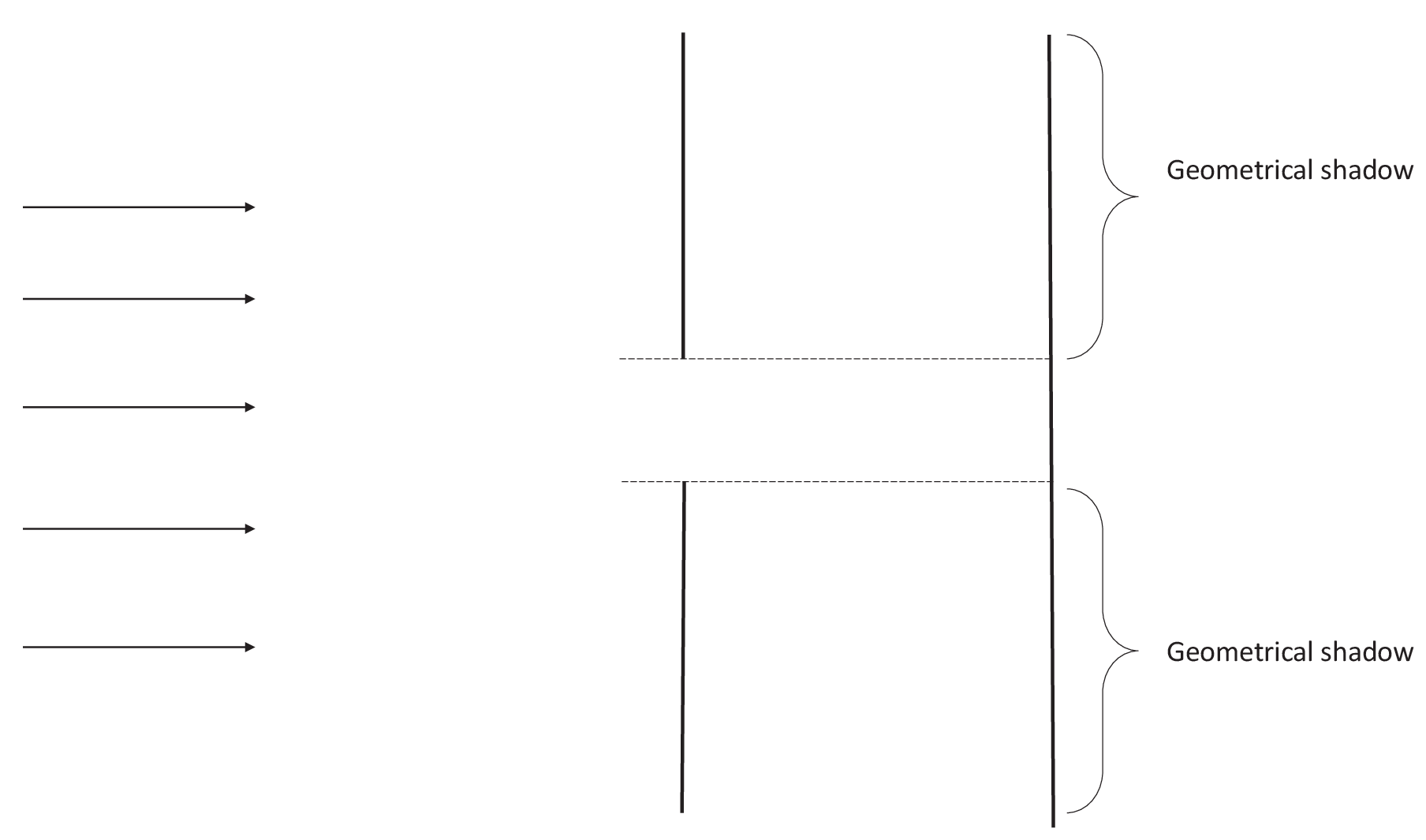}
\caption{A parallel light beam, which passes through an aperture. Newton thought that light never penetrates the geometrical shadow (he was wrong about that).}
\label{fig1}
\end{figure}

Even more, it is not very difficult to explain Snell's laws of refraction as well as the law of reflection of light using Newton's law of motion. Indeed, let us imagine particles of light passing through a medium 1 to medium 2, as shown in FIG.2. 
\begin{figure}[tb]
\includegraphics[width= 0.8\columnwidth]{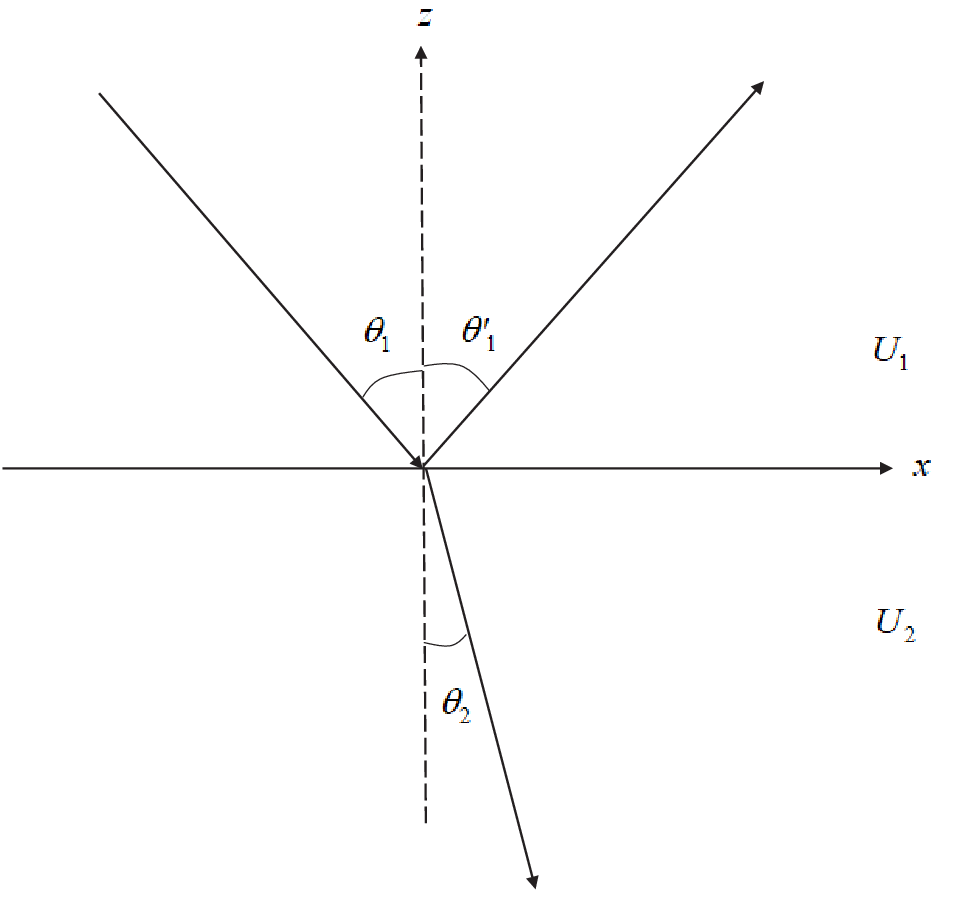}
\caption{Reflection and refraction of light.}
\label{fig2}
\end{figure}

At the boundary we assume that a force $\textbf{F}$ acts on the light corpuscles. We also assume that the force has a potential $U$. In the first medium the potential is $U_{1}=\text{const}.$, while in the second medium it is $U_{2}=\text{const.}$, and $U_{1}\neq U_{2}$. Then, since the potential energy $U$ changes in the $z$ direction only, the force $\textbf{F}$ also is in the $z$ direction. Then the momentum $P_{x}$ in the $x$ direction is conserved for both reflected and refracted light particles. If we apply the law of conservation of energy and of $P_{x}$ for the reflected light particles we have:
\begin{equation}
\frac{1}{2}m\textbf{v}_{1}^{2}+U_{1}=\frac{1}{2}m\textbf{v}_{1}^{\prime 2}+U_{1},\label{en1}
\end{equation}
where $m$ is the mass of the hypothetical light particles, and $\textbf{v}_{1}$ and $\textbf{v}_{1}^{\prime}$ are the velocities of incident and reflected light particles. It follows that the magnitude of these velocities are the same, i.e. $v_{1}=v_{1}^{\prime}$. The law of conservation of $P_{x}$ leads to:
\begin{equation}
mv_{1}\sin{\theta_{1}}=mv_{1}^{\prime}\sin{\theta_{1}^{\prime}}\label{Pz1}
\end{equation}
From Eqs. \eqref{en1} and \eqref{Pz1} we have that $\sin{\theta_{1}}=\sin{\theta_{1}^{\prime}}$ and thus $\theta_{1}=\theta_{1}^{\prime}$. This is the familiar law of reflection of light. 

Now, let us consider Snell's law of refraction. The law of conservation of energy leads to:
\begin{equation}
\frac{1}{2}m\textbf{v}_{1}^{2}+U_{1}=\frac{1}{2}m\textbf{v}_{2}^{2}+U_{2}\label{en2}
\end{equation}
Here $\textbf{v}_{2}$ is the velocity of refracted light particles. Now, obviously from Eq. \eqref{en2} we see that $v_{1}\neq v_{2}$. From the law of conservation of $P_{x}$ we have that:
\begin{equation}
v_{1}\sin{\theta_{1}}=v_{2}\sin{\theta_{2}},
\end{equation}
from which we obtain:
\begin{equation}
\dfrac{\sin{\theta_{1}}}{\sin{\theta_{2}}}=\frac{v_{2}}{v_{1}}\label{Snell1}
\end{equation}
The ratio $\sin{\theta_{1}}/\sin{\theta_{2}}$ depends only on the nature of the two media, as is Snell's law. 

We have now two powerful reasons in defense of the corpuscular theory of light - rectilinear motion of light (no penetration of light into the geometrical shadow) as well as simple explanation of the laws of reflection and refraction. 

We shall now see how the wave theory can account for these facts.

\subsection{Light as vibrations in the aether}
Hook \cite{Whittaker1989} considered light as vibrations of the aether. He considered a single pulse $AB$ that falls on the boundary between two media, as shown in FIG. 3. This vibration may be anything - it may be displacement of the aether particles, or may be contractions of its density, etc. People did not specify what kind of wave in the aether it was.
\begin{figure}[tb]
\includegraphics[width= 1.0\columnwidth]{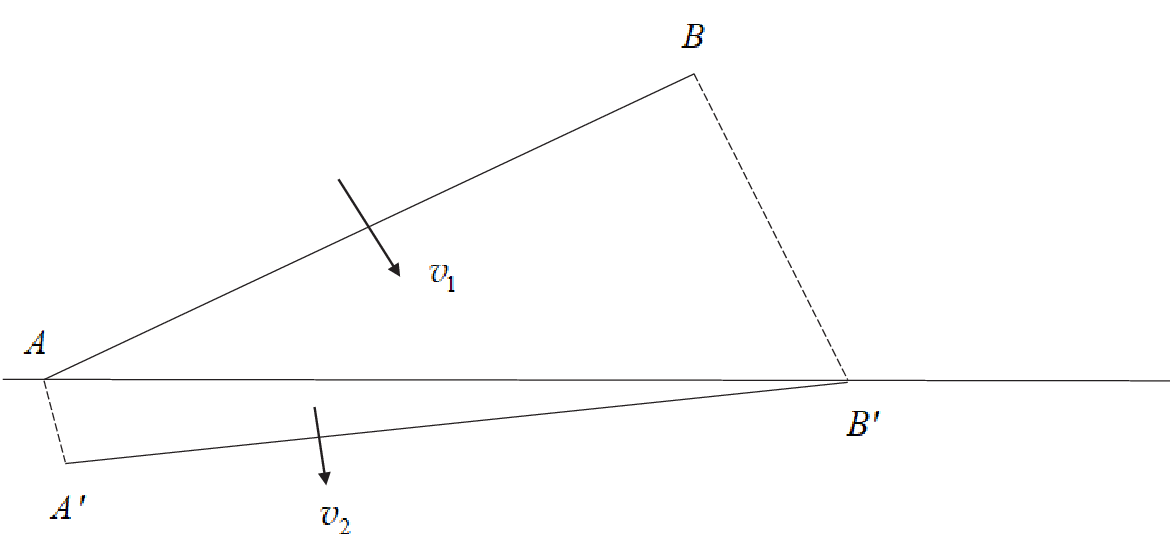}
\caption{A wave front, which is incident on the boundary between two media. The velocities of light in the two media are $v_{1}$ and $v_{2}$. According to Hook, if $v_{2}<v_{1}$, the point $A$ will move a shorter distance than point $B$ for the same amount of time, i.e. $|AA^{\prime}|<|BB^{\prime}|$. Then, the new wave front in the second medium will be inclined in a different angle than the angle of incidence.}
\label{fig3}
\end{figure}
Now, if $v_{2}<v_{1}$, the point $A$ will move a shorter distance than point $B$ for the same amount of time, i.e. $|AA^{\prime}|<|BB^{\prime}|$. Then, the new wave front in the second medium will be inclined in a different angle than the angle of incidence. Hook did not derive any equations but his account gives a hope that wave theory of light might be able to describe refraction of light. Note however that in Hook's theory, if $v_{2}<v_{1}$, the refracted light will be inclined toward the normal to the dividing boundary. This result is opposite to what Eq. \eqref{Snell1} would tell us.

Huygens established the principle that bears his name. This principle allows for the wave theory of light to answer all of Newton's objections \cite{Whittaker1989} and even to derive the laws of reflection and refraction. 

Indeed, it is well known that, according to Huygens' principle each point of the aether is a source of secondary spherical waves and the new wave front is just the envelope of these secondary wavelets.
\begin{figure}[tb]
\includegraphics[width= 0.1\columnwidth]{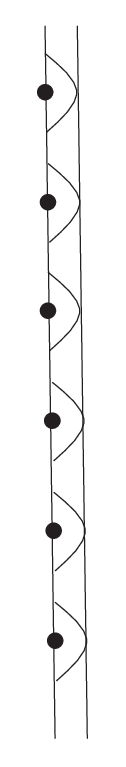}
\caption{Huygens' principle applied for a plane wave front. Each point of the plane is a point source for secondary wavelets. Their envelope gives the new wave front, which is also a plane.}
\label{fig4}
\end{figure}
In FIG. 4 we see how each point of a plane wave impulse creates secondary wavelets and the new wave front is again a plane. But Huygens' principle tells us much more. For instance, let us examine a pulse that passes through an aperture, as shown in FIG. 5.
\begin{figure}[tb]
\includegraphics[width= 1.0\columnwidth]{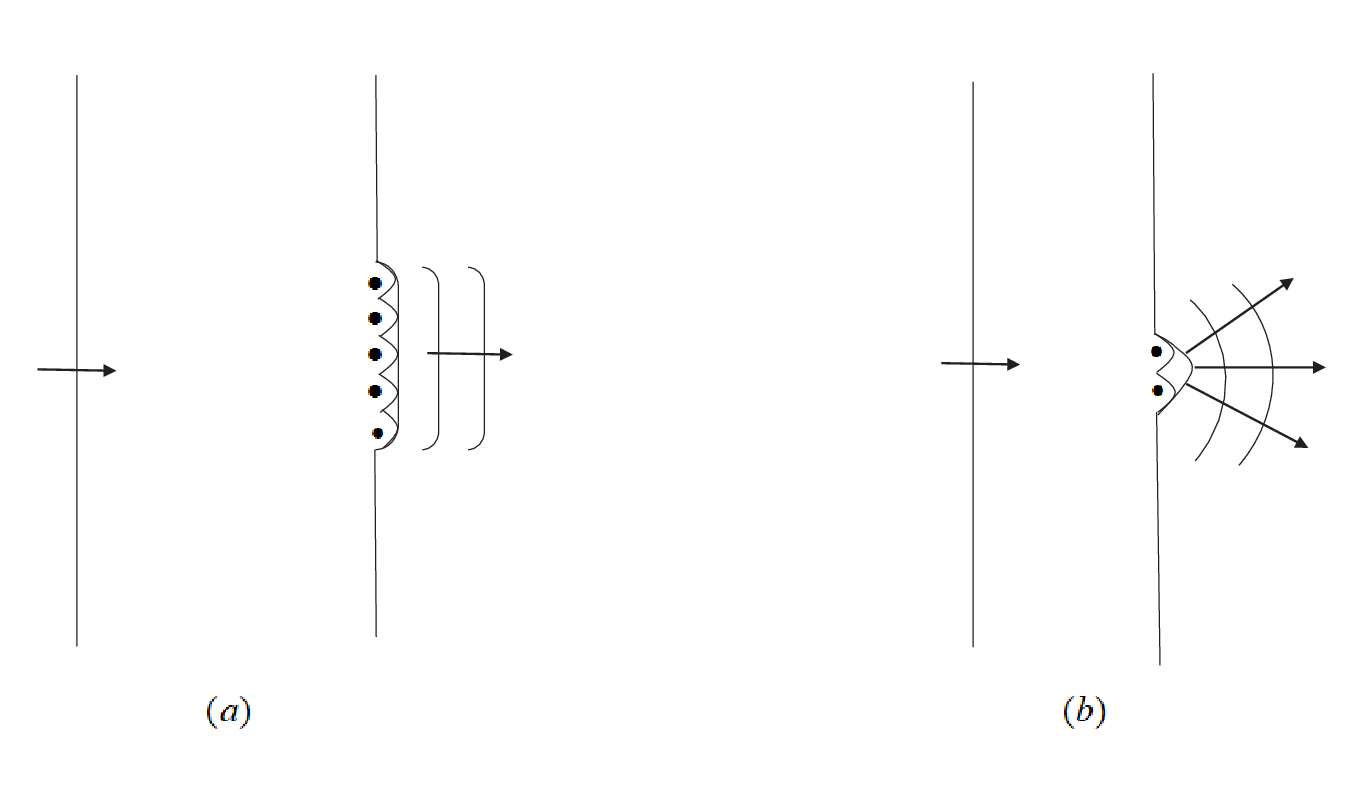}
\caption{Huygens's principle for a parallel beam of light that passes through an aperture. (a) If the aperture is large, the plane wave front remains almost a plane, and light does not penetrate into the geometrical shadow. (b) For a small aperture, the wave front \textit{after} the aperture is almost a circle, and light \textit{will} penetrate into the geometrical shadow (diffraction).}
\label{fig5}
\end{figure}
If the wave falls on a larger aperture, using Huygens' principle the new wave front is still almost a plane. However, if the aperture is very small, the new wave front is almost a circle and the light no longer moves in a rectilinear way. This phenomena, in which waves penetrate into the geometrical shadow by deviating from rectilinear motion is called \textit{diffraction}. Now, finally Huygens can explain why light can move in a rectilinear way, as if a beam of particles. Perhaps ordinary apertures, through which light usually passes are just too large, as in Fig. 5 (a). Huygens principle tells us that if we have smaller apertures, light will penetrate into the geometrical shadow. Huygens' principle cannot tell us how small apertures should be but nonetheless it predicts the phenomena of diffraction for small apertures. But more importantly Huygens' principle can in principle explain why light waves move in a rectilinear way and for large apertures do not penetrate into the geometrical shadow.

Huygens' principle not only answers Newton's objection and can explain the rectilinear motion of light but it can also derive the laws of reflection and refraction.

Indeed, let us examine a wave impulse that falls on the boundary between two media, as shown in FIG. 6 and FIG. 7.

We assume that the speeds of light $v_{1}$ and $v_{2}$ in these two media are different, i.e. $v_{1}\neq v_{2}$. This may be explained by the fact that perhaps the aether's density or the ather's elasticity is different in the two media. Whatever the explanation, it is quite obvious that one can easily imagine why $v_{1}\neq v_{2}$. In addition, we also assume that without loss of generality, $v_{1,2}\neq c$, where $c$ is the speed of light in vacuum.

\begin{figure}[tb]
\includegraphics[width= 1.0\columnwidth]{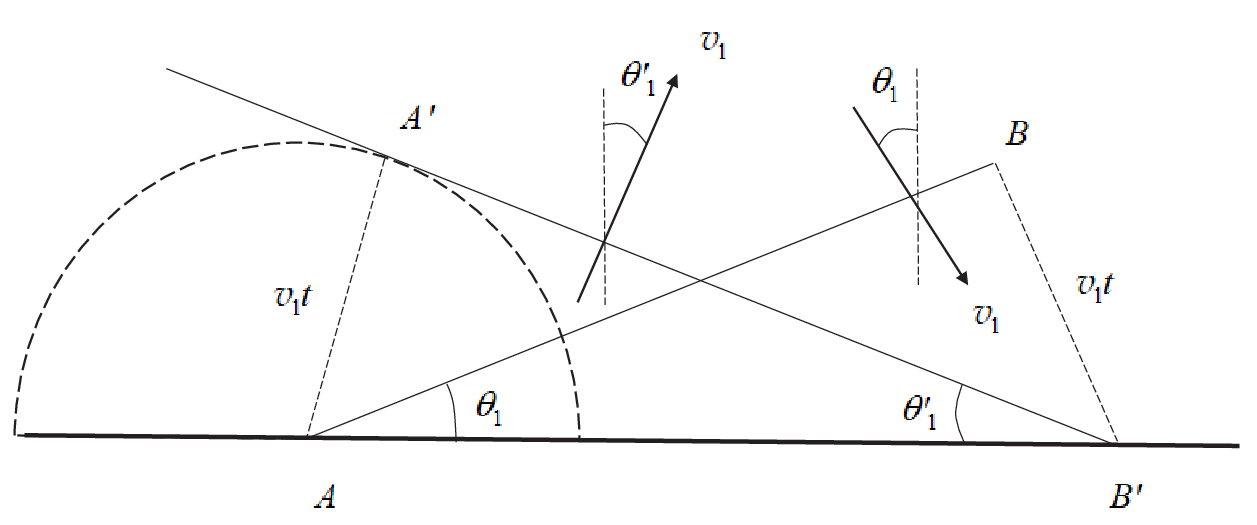}
\caption{Reflection of light according to Hyygen's principle. As the point $A$ reaches the boundary between the two media, at point $A$ a spherical wave is emitted (according to Huygens' principle). After some time $t$ point $B$ reaches the boundary at point $B^{\prime}$. Obviously, after this time, $BB^{\prime}=v_{1}t$, while the radius of the sphere becomes $AA^{\prime}=v_{1}t$. It is obvious then that $AA^{\prime}=BB^{\prime}=v_{1}t$. Then, the rectangular triangles $AA^{\prime}B^{\prime}$ and $ABB^{\prime}$ are congruent and then $\theta_{1}=\theta_{1}^{\prime}$}
\label{fig6}
\end{figure}

We start with the law of reflection (FIG. 6): as the point $A$ reaches the boundary between the two media, at point $A$ a spherical wave is emitted (according to Huygens' principle). After some time $t$ point $B$ reaches the boundary at point $B^{\prime}$. Obviously, after this time, $BB^{\prime}=v_{1}t$, while the radius of the sphere becomes $AA^{\prime}=v_{1}t$. It is obvious then that $AA^{\prime}=BB^{\prime}=v_{1}t$. Then, the rectangular triangles $AA^{\prime}B^{\prime}$ and $ABB^{\prime}$ are congruent and then $\theta_{1}=\theta_{1}^{\prime}$. On the other hand, the angles $\theta_{1}$ and $\theta_{1}^{\prime}$ are the angles of incidence and reflection. Then we have derived the law of reflection, that is the equality of the angles of incidence and of reflection.

If we use Huygens' principle we can also prove Snell's law of refraction. Indeed, let us examine a light pulse that passes through a medium 1 to a medium 2, as shown in FIG. 7.

\begin{figure}[tb]
\includegraphics[width= 1.1\columnwidth]{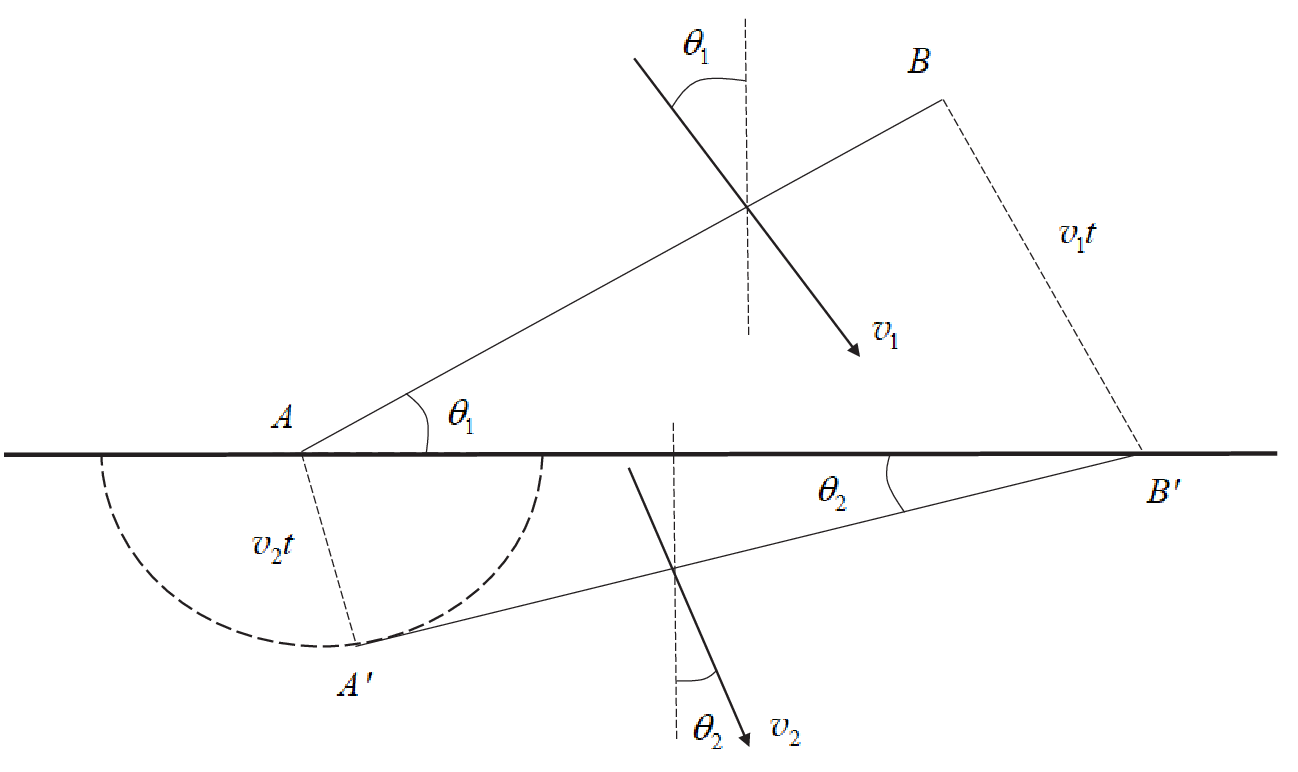}
\caption{Refraction of light according to Huygen's principle. As point $A$ reaches the boundary, a spherical wave is emitted. After some time $t$ point $B$ reaches the boundary and the distance it has traveled is $BB^{\prime}=v_{1}t$. For the same amount of time, the radius of the spherical wave with center $A$ has become $AA^{\prime}=v_{2}t$. On the other hand we have: $\sin{\theta_{1}}=BB^{\prime}/AB^{\prime}=v_{1}t/AB^{\prime}$ as well as: $\sin{\theta_{2}}=AA^{\prime}/AB^{\prime}=v_{2}t/AB^{\prime}$. We divide these two equations and we derive Snell's law: $\sin{\theta_{1}}/\sin{\theta_{2}}=v_{1}/v_{2}$.}
\label{fig7}
\end{figure}

As point $A$ reaches the boundary, a spherical wave is emitted. After some time $t$ point $B$ reaches the boundary and the distance it has traveled is $BB^{\prime}=v_{1}t$. For the same amount of time, the radius of the spherical wave with center $A$ has become $AA^{\prime}=v_{2}t$. Now, obviously
\begin{equation}
\sin{\theta_{1}}=\frac{BB^{\prime}}{AB^{\prime}}=\frac{v_{1}t}{AB^{\prime}}\label{q1}
\end{equation}
On the other hand
\begin{equation}
\sin{\theta_{2}}=\frac{AA^{\prime}}{AB^{\prime}}=\frac{v_{2}t}{AB^{\prime}}\label{q2}
\end{equation}

We divide both Eqs. \eqref{q1} and \eqref{q2} and we obtain:
\begin{equation}
\frac{\sin{\theta_{1}}}{\sin{\theta_{2}}}=\dfrac{v_{1}}{v_{2}}.\label{Snell2}
\end{equation}
The ratio $\sin{\theta_{1}}/\sin{\theta_{2}}$ depends only on the nature of the media, i.e. on the velocities $v_{1}$ and $v_{2}$ which are determined by the type of media only. This is just Snell's law. Note that Eq. \eqref{Snell2} is quite the opposite of Eq. \eqref{Snell1}. 

If we define the indexes of refraction: $n_{1}=c/v_{1}$ and $n_{2}=c/v_{2}$, where $c$ is the speed of light in vacuum, then Eq. \eqref{Snell2} becomes:
\begin{equation}
\frac{\sin{\theta_{1}}}{\sin{\theta_{2}}}=\dfrac{n_{2}}{n_{1}}.
\end{equation}
This is the familiar way Snell's law is written in textbooks.

In other words, the wave theory of light can also explain reflection and refraction.

Finally, due to the works of Young and Fresnel on the interference and diffraction of light, as well as the successful explanation of the phenomena of double refraction and polarization, wave theory finally triumphed in the 19th century \cite{Whittaker1989}.

\section{How to describe waves}
\subsection{Plane waves}

Huygens' principle was just a heuristic device to try to make intuitive sense of how waves propagate. Nowadays we know how waves propagate, since the mathematics of waves is way better understood.

Let us consider some function $f(x)$ (it may be pressure distribution, or density distribution, or component of electric or magnetic field, etc.), that moves with some speed $c$.

\begin{figure}[tb]
\includegraphics[width= 1.0\columnwidth]{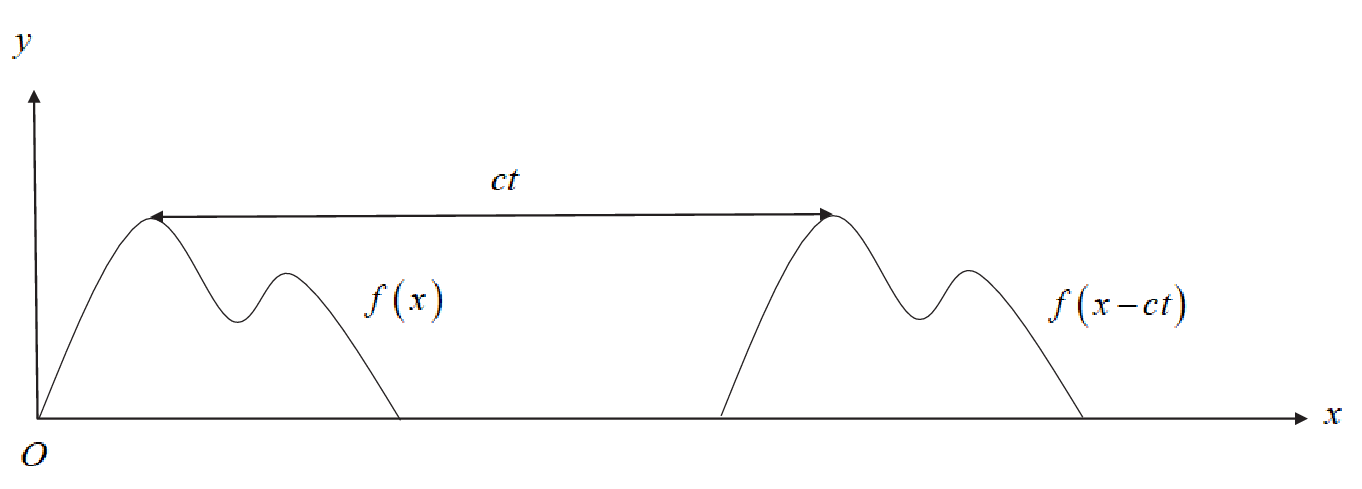}
\caption{Motion of a wave described by a function $f(x)$. After time $t$, the function is translated to $f(x-ct)$.}
\label{fig8}
\end{figure}

Then, after some time $t$, the function is translated to a distance $ct$ (FIG. 8). The translated function becomes $f(x-ct)$. The most familiar wave is the cosine function: $f(x)=A\cos{(kx)}$, which after a distance $\lambda=2\pi/k$ repeats itself. This $\lambda$ is the familiar wavelength. After time $t$ the wave becomes:
\begin{equation}
f(x-ct)=A\cos{\left[ k(x-ct)\right]}=A\cos{(kx-kct)}.
\end{equation}
If we substitute
\begin{equation}
\omega=kc
\end{equation}
we have for the moving cosine:
\begin{equation}
f(x-ct)=A\cos{(kx-\omega t)}\label{pw1}
\end{equation}
The meaning of $\omega$ is obvious. Let us fix our attention at some particular point with a coordinate $x$. As the wave propagates to the right, the function $f$ repeats itself at this particular position $x$ after a time $T=2\pi/\omega$. Then $T$ is the familiar period of the wave, and that is why $\omega$ is called frequency of the wave. In other words $\omega$ tells us how frequently the function $f$ repeats itself at some fixed point $x$. The argument $kx-\omega t$ is called phase of the wave. Since the function cosine is periodic, points with phase difference $2\pi n$ for any arbitrary integer $n=0,\pm 1, \pm 2,...$ are disturbed in the same way, since the function $f$ has the same value for these points.

Let us examine the function $f(x-ct)=A\cos{\left(kx-\omega t\right)}$. At time $t$ and position $x$ the phase is $kx-\omega t$. What is the new coordinate $x+\delta x$ of the point, which will have the same phase at the next moment $t+\delta t$. It is easy to determine:
\begin{equation}
k.(x+\delta x)-\omega .(t+\delta t)=kx-\omega t
\end{equation}
From this equation we have:
\begin{equation}
k\delta x=\omega\delta t
\end{equation}
In other words, the point with a fixed phase, moves with a velocity (phase velocity):
\begin{equation}
v_{\text{ph}}=\frac{\delta x}{\delta t}=\frac{\omega}{k}\equiv c.
\end{equation}

In a similar way, we can examine a 3D wave:
\begin{equation}
f(x,y,z,t)=A\cos{\left(k_{1}x+k_{2}y+k_{3}z-\omega t\right)}\equiv A\cos{\left(\textbf{k}\cdot \textbf{r}-\omega t\right)}\label{planewave}
\end{equation}
We have introduced the wave vector: $\textbf{k}=(k_{1},k_{2},k_{3})$ and the position vector $\textbf{r}=(x,y,z)$. It is easy to find all points with the same phase at some fixed moment $t$:
\begin{equation}
k_{1}x+k_{2}y+k_{3}z-\omega t=\text{const}.
\end{equation}
But since $t$ is fixed, i.e. $t=\text{const}.$, we have that
\begin{equation}
k_{1}x+k_{2}y+k_{3}z=\text{const}.\label{plane}
\end{equation}
But this is an equation of a plane. Recall that $Ax+By+Cz=\text{const}.$ is an equation of a plane, with a normal vector $\textbf{N}=(A,B,C)$. Similarly Eq. \eqref{plane} tells us that all points with the same phase (the so called phase surface, or wave front) lie in a plane, with a normal vector $\textbf{k}$, as shown in FIG. 9. That is why the wave, described by Eq. \eqref{planewave} is called a plane wave.

\begin{figure}[tb]
\includegraphics[width= 1.0\columnwidth]{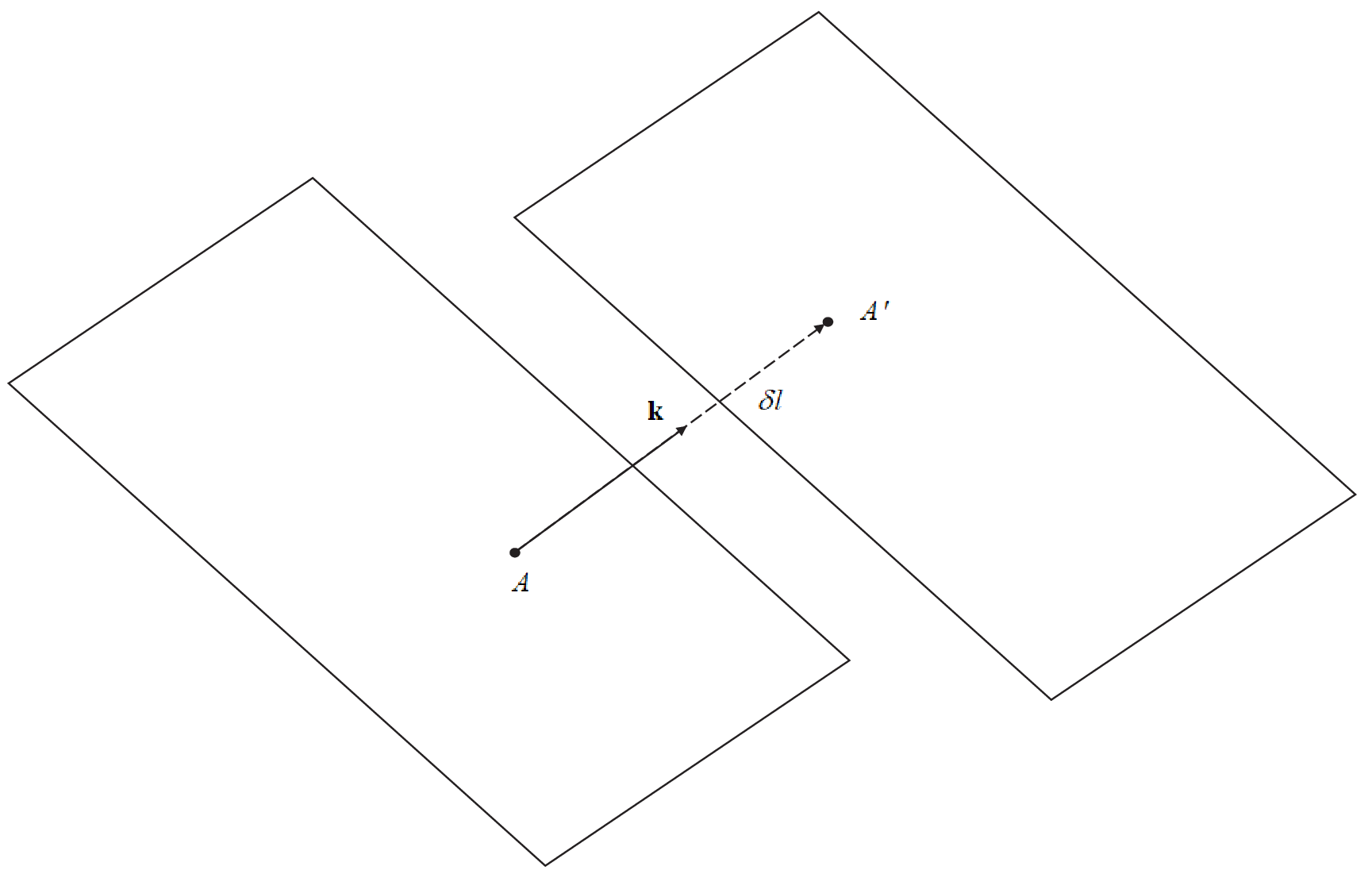}
\caption{Two wave fronts of a plane wave. The wave vector $\textbf{k}$ is perpendicular to the planes. After a time interval $\delta t$ the lower plane moves a distance $\delta l$ to the new position.}
\label{fig9}
\end{figure}

Let us examine how the phase surface (the plane) moves. After a time interval $\delta t$, the plane with the same phase has moved to a new position, as shown in FIG. 9. Let us translate point $A$ along the vector $\textbf{k}$ to the new plane in point $A^{\prime}$ (see FIG.9). The length $\delta l$ is the distance $AA^{\prime}$. We can easily calculate the speed necessary for point $A$ to move to point $A^{\prime}$. We have that the phase is the same for both surfaces, i.e.
\begin{align}
& k_{1}.(x+\delta x)+k_{2}.(y+\delta t)+k_{3}.(z+\delta z)-\omega .(t+\delta t)=\notag\\
&k_{1}x+k_{2}y+k_{3}z-\omega t
\end{align} 
Simplifying we have:
\begin{equation}
k_{1}\delta x+k_{2}\delta y+k_{3}\delta z=\omega\delta t
\end{equation}
But we can group $\delta x$, $\delta y$ and $\delta z$ into a vector $\delta \textbf{l}$ and this equation is transformed to:
\begin{equation}
\textbf{k}\cdot\delta \textbf{l}=\omega\delta t
\end{equation}
But $\delta \textbf{l}\parallel \textbf{k}$ then $\textbf{k}\cdot\delta \textbf{l} =k\delta l$. Then we have
\begin{equation}
k\delta l=\omega\delta t
\end{equation}
Then we can find the phase velocity
\begin{equation}
v_{\text{ph}}=\frac{\delta l}{\delta t}=\frac{\omega}{k}\equiv c.
\end{equation}
If the wave front moves in a medium with index of refraction $n$, then the same reasoning gives:
\begin{equation}
v_{\text{ph}}=\frac{\delta l}{\delta t}=\frac{\omega}{k}\equiv \frac{c}{n}.
\end{equation}

\subsection{Arbitrary wave}
Let us consider now an arbitrary function, with a fixed frequency:
\begin{equation}
f(\textbf{r},t)=A(\textbf{r})\cos{\left[ S(\textbf{r})-\omega t\right]}\label{arbw}
\end{equation}
Here $S(\textbf{r})$ is just an arbitrary phase of the function $f$. For a plane wave, $\textbf{S}=\textbf{k}\cdot \textbf{r}$. But in the genral case \eqref{arbw} $S$ can be quite arbitrary.

The function $f$ is still periodic in time with a period $T=2\pi/\omega$. Now, the question is: when this function is a wave? For plane waves it was very easy to discover the condition, when the functions \eqref{pw1} and \eqref{planewave} are actually waves. It was necessary that $\omega/k=c/n$. But it is more difficult to establish such a condition for arbitrary waves, like \eqref{arbw}. Fortunately we know the correct way to discover whether a function is a wave. It has to obey the wave equation (this is our \textit{definition} of a wave):
\begin{equation}
\frac{\partial^{2}f}{\partial x^{2}}+\frac{\partial^{2}f}{\partial y^{2}}+\frac{\partial^{2}f}{\partial z^{2}}-\frac{1}{v^{2}}\frac{\partial^{2}f}{\partial t^{2}}=0
\end{equation}
where $v=c/n$ is the speed of light in the medium with index of refraction $n$. If we substitute $v=c/n$ in the above equation we have:
\begin{equation}
\frac{\partial^{2}f}{\partial x^{2}}+\frac{\partial^{2}f}{\partial y^{2}}+\frac{\partial^{2}f}{\partial z^{2}}-\frac{n^{2}}{c^{2}}\frac{\partial^{2}f}{\partial t^{2}}=0 \label{we}
\end{equation}
Now, it is very easy to confirm the condition when the simple functions \eqref{pw1} and \eqref{planewave} are really waves in the particular medium. If we substitute \eqref{pw1} and \eqref{planewave} into the above wave equation we see that the condition for waves is: $\omega/k=c/n$, as expected. Then in order to discover when the more complicated \eqref{arbw} is a wave, we also have to substitute it into the wave equation \eqref{we}.

Before we do that, let us calculate the phase velocity $v_{\text{ph}}$ of the arbitrary wave \eqref{arbw}. Let us consider all points with the same phase, i.e. $S(\textbf{r})-\omega t=\text{const}$. When we select a specific moment of time $t$, i.e. $t=\text{const}.$, then all points with the same phase are such that: $S(\textbf{r})=\text{const}$. This is an equation of arbitrary surface, not necessarily a plane, as shown in Fig. 10.

\begin{figure}[tb]
\includegraphics[width= 1.0\columnwidth]{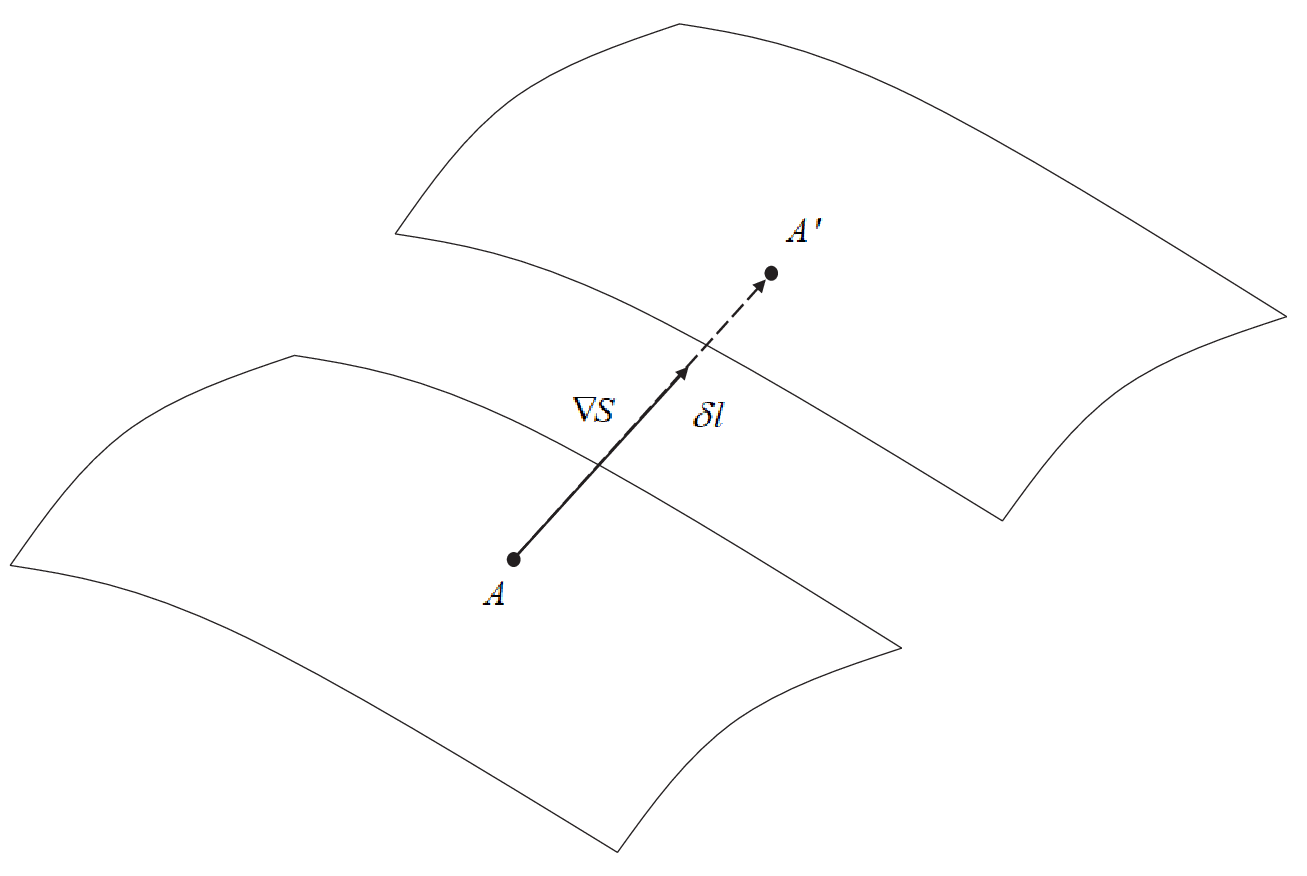}
\caption{Two wave fronts of an arbitrary wave. After a time interval $\delta t$, the lower surface $S(\textbf{r})=\text{const}.$ moves to the upper in a distance $\delta l$. The vector $\nabla S$ is perpendicular to the wave fronts. Point $A$ has moved to point $A^{\prime}$ along $\nabla S  $.}
\label{fig10}
\end{figure}

After time $\delta t$, the phase surface has moved. We can easily find the phase velocity. Indeed, we have that the two adjacent surfaces have the same phase if:
\begin{equation}
S(x,y,z)-\omega t=S(x+\delta x,y+\delta y,z+\delta z)-\omega(t+\delta t).
\end{equation}
We Taylor expand $S(x+\delta x,y+\delta y,z+\delta z)$ and we have:
\begin{equation}
S(x,y,z)-\omega t = S(x,y,z)+\frac{\partial S}{\partial x}\delta x+\frac{\partial S}{\partial x}\delta x+\frac{\partial S}{\partial x}\delta x-\omega t-\omega\delta t.
\end{equation}
(we do not use approximate sign, since we consider infinitesimal quantities). If we group $\delta x$, $\delta y$ and $\delta z$ into a vector $\delta \textbf{l}=(\delta x, \delta y, \delta z)$ and if we use the gradient of $S$: $\nabla S=(\frac{\partial S}{\partial x},\frac{\partial S}{\partial y},\frac{\partial S}{\partial z})$, we can reduce the above equation to:
\begin{equation}
\nabla S\cdot\delta \textbf{l}=\omega\delta t\label{vph1}
\end{equation}

Now, it is well known that $\nabla S$ is normal to the plane \cite{Boas2005, Nearing2010}. If we select points $A$ and $A^{\prime}$ in such a way, that $\delta \textbf{l}$ is also normal to the plane, then $\delta \textbf{l}\parallel\nabla S$. Then $\nabla S\cdot\delta \textbf{l}=|\nabla S|\delta l$ and Eq. \eqref{vph1} becomes:
\begin{equation}
|\nabla S|\delta l=\omega\delta t
\end{equation}
From this equation we can finally calculate the phase velocity:
\begin{equation}
v_{\text{ph}}=\frac{\delta l}{\delta t}=\frac{\omega}{|\nabla S|}.\label{go0}
\end{equation}
One might expect that the condition for the function \eqref{arbw} to be wave is that
\begin{equation}
v_{\text{ph}}\equiv\frac{\omega}{|\nabla S|}=\frac{c}{n}\label{go}
\end{equation}
But this is wrong! We shall see that condition \eqref{go} is \textit{not} a condition for $f$ to be a wave. It is a condition for something \textit{else} entirely. Indeed, in order for \eqref{arbw} to be a wave, it has to obey the wave equation \eqref{we}. That is the \textit{true} condition. We shall  later substitute \eqref{arbw} into the wave equation \eqref{we} and we shall see that there will be some derivatives of the amplitude $A(\textbf{r})$ as well as second derivatives of $S$. No such derivatives are seen in Eq. \eqref{go}. Thus we see that the simple condition \eqref{go} will \textit{not} suffice as in the case of plane waves! Condition \eqref{go} is not a condition for $f$ to be a wave, but for something else.

\textit{We shall prove that condition} \eqref{go} \textit{is a condition for the wave to behave as a beam of particles}! \textit{Indeed, if \eqref{go} is true, and if the index of refraction $n=\text{const}.$ (which is the most frequent case), the phase points of a plane wave will move in \textit{straight} lines (rays), since when $n=\text{const}$, then \eqref{go} implies that $v_{\text{ph}}=\text{const}$. In this way the phase points will move in straight lines and will never penetrate into the geometrical shadow}. 

Usually people prefer to square \eqref{go} and we have
\begin{equation}
\left(\nabla S\right)^{2}=\frac{n^{2}\omega^{2}}{c^{2}}.\label{eikonal}
\end{equation}
This is the very famous Eikonal equation. This is the condition for waves to behave like a beam of particles. This regime, when light obeys the Eikonal equation is called geometrical optics. People usually use Eq. \eqref{eikonal} to find the rays in the case when $n=n(x,y,z)$ change from point to point. This shows that even in geometrical optics, light \textit{can} penetrate into the geometrical shadow, so long as $n$ is a function of $x$, $y$ and $z$. However $n=\text{const}.$ is almost always true in everyday experience. 

Now, what is left is to check if it is possible for the function \eqref{arbw} to actually obey the Eikonal equation \eqref{eikonal}. To this end we shall substitute \eqref{arbw} into the wave equation \eqref{we}. But first, in order to make the calculation simple, we shall use complex notation:
\begin{equation}
f(\textbf{r},t)=A(\textbf{r})\cos{\left[ S(\textbf{r})-\omega t\right]}=\text{Re}(F)
\end{equation}
where $F$ is a complex wave:
\begin{equation}
F=A(\textbf{r})e^{iS(\textbf{r})}e^{-i\omega t}
\end{equation}
We shall rewrite the wave equation into:
\begin{equation}
\nabla^{2}f-\frac{n^{2}}{c^{2}}\frac{\partial ^{2}f}{\partial t^{2}}=0\label{waveeqn}
\end{equation}
We have that:
\begin{align}
&\nabla^{2}f=\text{Re}\left[\nabla\cdot\left(e^{iS}e^{-i\omega t}\nabla A+iAe^{iS}e^{-i\omega t}\nabla S\right)\right]\notag\\
&=\text{Re}\left[\left(\nabla^{2}A+2i\nabla A\cdot\nabla S-A\left(\nabla S\right)^{2}+iA\nabla^{2}S\right)e^{iS}e^{-i\omega t}\right]\label{Laplf}
\end{align}
Now, let us assume that for some reasons (which we shall investigate below), the third term is much larger than the others. Then we have
\begin{equation}
\nabla^{2}f\approx\text{Re}\left[-A\left(\nabla S\right)^{2}e^{iS}e^{-i\omega t}\right]\label{Lalplf2}
\end{equation}
In this special regime the wave equation becomes:
\begin{align}
\text{Re}\left[\left(-A\left(\nabla S\right)^{2}+\frac{n^{2}\omega^{2}}{c^{2}}A\right)e^{iS}e^{-i\omega t}\right]=0
\end{align}
Simplifying:
\begin{equation}
A\cos{(S-\omega t)}\left(\frac{n^{2}\omega^{2}}{c^{2}}-\left(\nabla S\right)^{2}\right)=0.
\end{equation}
When we cancel $A\cos{(S-\omega t)}$, we finally obtain the Eikonal equation:
\begin{equation}
\frac{n^{2}\omega^{2}}{c^{2}}-\left(\nabla S\right)^{2}=0.
\end{equation}
But if the Eikonal equation is true, a wave can behave as a beam of particles. And we now finally see the conditions for that -- it must be the case that the term $\left(\nabla S\right)^{2}$ in Eq. \eqref{Laplf} as well as the term $n^{2}\omega^{2}/c^{2}$ should be much larger than all other terms in the wave equation. What is left then is to find out what is the physical meaning of these terms being the largest.

Imagine we have two very close wave fronts, such that the phase difference between them is $2\pi$, similar to FIG. 10. However, for these two surfaces we shall call the distance $\delta l=\lambda$, and it will correspond to the wavelength of this wave. Then we have:
\begin{equation}
S(x+\delta x, y+\delta y,z+\delta z)-S(x,y,z)=2\pi
\end{equation}
Taylor expanding we have:
\begin{equation}
\frac{\partial S}{\partial x}\delta x+\frac{\partial S}{\partial y}\delta y+\frac{\partial S}{\partial z}\delta z=2\pi
\end{equation}
Which is the same as:
\begin{equation}
\nabla S\cdot\delta \textbf{l}=2\pi
\end{equation}
But again we have that $\delta \textbf{l}\|\nabla S$ and $\delta l=\lambda$. Then we have:
\begin{equation}
|\nabla S|\lambda=2\pi
\end{equation}
In that case we finally see the meaning of $\nabla S$:
\begin{equation}
|\nabla S|=\frac{2\pi}{\lambda}
\end{equation}
$\nabla S$ simply corresponds to the wave vector $\textbf{k}$ for plane waves. The only difference is that this time the wave vector depends on $x$, $y$ and $z$, i.e. we have position dependent wave vector $\textbf{k}$. This will lead also to position dependent wavelength $\lambda=\lambda(\textbf{r})$. If $(\nabla S)^{2}$ is much greater than all other terms in Eq. \eqref{Laplf}, it means that the wavelength is quite small. In this way we reach the conclusion that for sufficiently short wavelengths, we have the regime of geometrical optics and no penetration in the geometrical shadow (if the index of refraction $n=\text{const}$).

The term $n\omega/c=\omega/v=2\pi/\lambda_{0}$, where $\lambda_{0}$ is the distance the wave has traveled in one period. People are used to think that $\lambda=\lambda_{0}$, but this is true only for plane waves or in the regime of geometrical optics. Indeed, the Eikonal equation can in fact be rewritten as $\lambda=\lambda_{0}$. Just substitute $|\nabla S|=2\pi\lambda$ and $n\omega=2\pi/\lambda_{0}$ into the Eikonal equation Eq. \eqref{eikonal} (or \eqref{go}) and the result $\lambda=\lambda_{0}$ immediately follows.

One last point. It is easy to see that if the phase $S$ is sufficiently large (to be made more precise below), then this would effectively lead to the regime of geometrical optics. Indeed, let us substitute $S=\tilde{S}/\epsilon$. Here $\tilde{S}\sim 1$, while $\epsilon \ll 1$. In that case  $S$ is very large. If we substitute $S$ into Eq. \eqref{Laplf}, the terms in the brackets become:

\begin{align}
&\left(\nabla^{2}A+2i\epsilon^{-1}\nabla A\cdot\nabla\tilde{S}-A\epsilon^{-2}\left(\nabla S\right)^{2}+iA\epsilon^{-1}\nabla^{2}\tilde{S}\right)\times\notag\\
&\times e^{i\tilde{S}/\epsilon}e^{-i\omega t}\label{largephase}
\end{align}
The largest of the terms in the brackets is the term with $\epsilon^{-2}$. Then for large $S$ (small $\epsilon$) we have the regime of geometrical optics.

\section{Time-dependent Eikonal equation}
So far, we have considered an arbitrary wave with the same frequency $\omega$ in Eq. \eqref{arbw}. We will consider now even a wave, which has arbitrary time dependence:
\begin{equation}
f(\textbf{r},t)=A(\textbf{r})\cos{S(\textbf{r},t)}=\text{Re}\left(A(\textbf{r})e^{iS(\textbf{r},t)}\right)
\end{equation}
In order to calculate the phase velocity in this case, we take two points $A$ and $A^{\prime}$ (see FIG. 10), such that again $\delta \textbf{l}\parallel\nabla S$.
The two phases are the same, i.e.
\begin{equation}
S(x,y,z,t)=S(x+\delta x, y+\delta y, z+\delta z, t+\delta t)
\end{equation}
After Taylor expanding, we obtain:
\begin{equation}
\nabla S\cdot\delta \textbf{l}=\frac{\partial S}{\partial t}\delta t.
\end{equation}
Since $\delta \textbf{l}\parallel\nabla S$, we have that:
\begin{equation}
v_{\text{ph}}=\frac{\delta l}{\delta t}=\frac{\partial_{t}S}{|\nabla S|}.
\end{equation}

Again, in order for the wave front to move like a beam of particles, we expect that
\begin{equation}
v_{\text{ph}}=c/n
\end{equation}
Then, this reduces to:
\begin{equation}
\frac{\partial_{t}S}{|\nabla S|}=c/n
\end{equation}
When we square, we have:
\begin{equation}
(\nabla S)^{2}=\frac{n^{2}}{c^{2}}\left(\frac{\partial S}{\partial t}\right)^{2}\label{eikonal2}
\end{equation}
This is the new condition for geometrical optics.

It is not very difficult to prove that if $|\nabla S|$ and $\partial_{t}S$ are sufficiently large, the wave equation is reduced to the Eikonal equation \eqref{eikonal2}. Indeed, if $\lambda=2\pi/|\nabla S|$ is quite small, $\nabla^{2}f$ from Eq. \eqref{Laplf} is again reduced to \eqref{Lalplf2}. In addition:
\begin{equation}
\frac{\partial ^{2}f}{\partial t^{2}}=\text{Re}\left[Ae^{iS}\left(i\frac{\partial^{2}S}{\partial t^{2}}-\left(\frac{\partial S}{\partial t}\right)^{2}\right)\right]
\end{equation}
If the last term is the largest:
\begin{equation}
\frac{\partial ^{2}f}{\partial t^{2}}\approx\text{Re}\left[-Ae^{iS}\left(\frac{\partial S}{\partial t}\right)^{2}\right]
\end{equation}
Finally if we substitute this equation and Eq. \eqref{Lalplf2} into the wave equation \eqref{waveeqn}, we obtain again the Eikonal equation \eqref{eikonal2}. Therefore arbitrary waves with arbitrary time dependence may behave as a beam of particles.

Finally, the meaning of $\partial_{t}S$ is analogous to the frequency, but this time position and time dependent. Then $n\partial_{t}S/c=\lambda_{0}$ gives the distance the wave travels for one period at position $\textbf{r}=(x,y,z)$ at time $t$. Sufficiently small wavelengths $\lambda$ and $\lambda_{0}$ or sufficiently large $S$ will lead to the regime of geometrical optics and to Eq. \eqref{eikonal2}.

\section{Electron gun experiments}
In this section we shall present the electron gun and shall introduce the effects of interference of electrons. We shall follow \cite{Maudlin2019}. These are however hypothetical experiments. For real experiments see Refs. \cite{Frabboni2007} and \cite{Frabboni2008}.

Each battery has a positive and a negative end. These ends are called electrodes. The positive electrode is called an anode, while the negative is called a cathode. Let us connect the two ends of the battery with wires to two metal plates (see FIG. 11). Also, let us make aperture in the plate, connected to the anode, and behind the anode we place phosphor-coated screen.

\textit{Experiment 1}: When the cathode is very hot (due to the heater), we see on the screen a bright spot. However, if we turn down the heater, the spot gets dimmer and dimmer. If we continue to turn down the heater even more, instead of a spot, individual flashes appear on the screen, in exactly the same place, where the spot was (FIG. 11). If we continue to turn down the heater even further, the flashes appear one at a time with a great interval of time between them. The individual flashes are obviously particles. These are the familiar \textit{electrons}.

\begin{figure}[tb]
\includegraphics[width= 1.0\columnwidth]{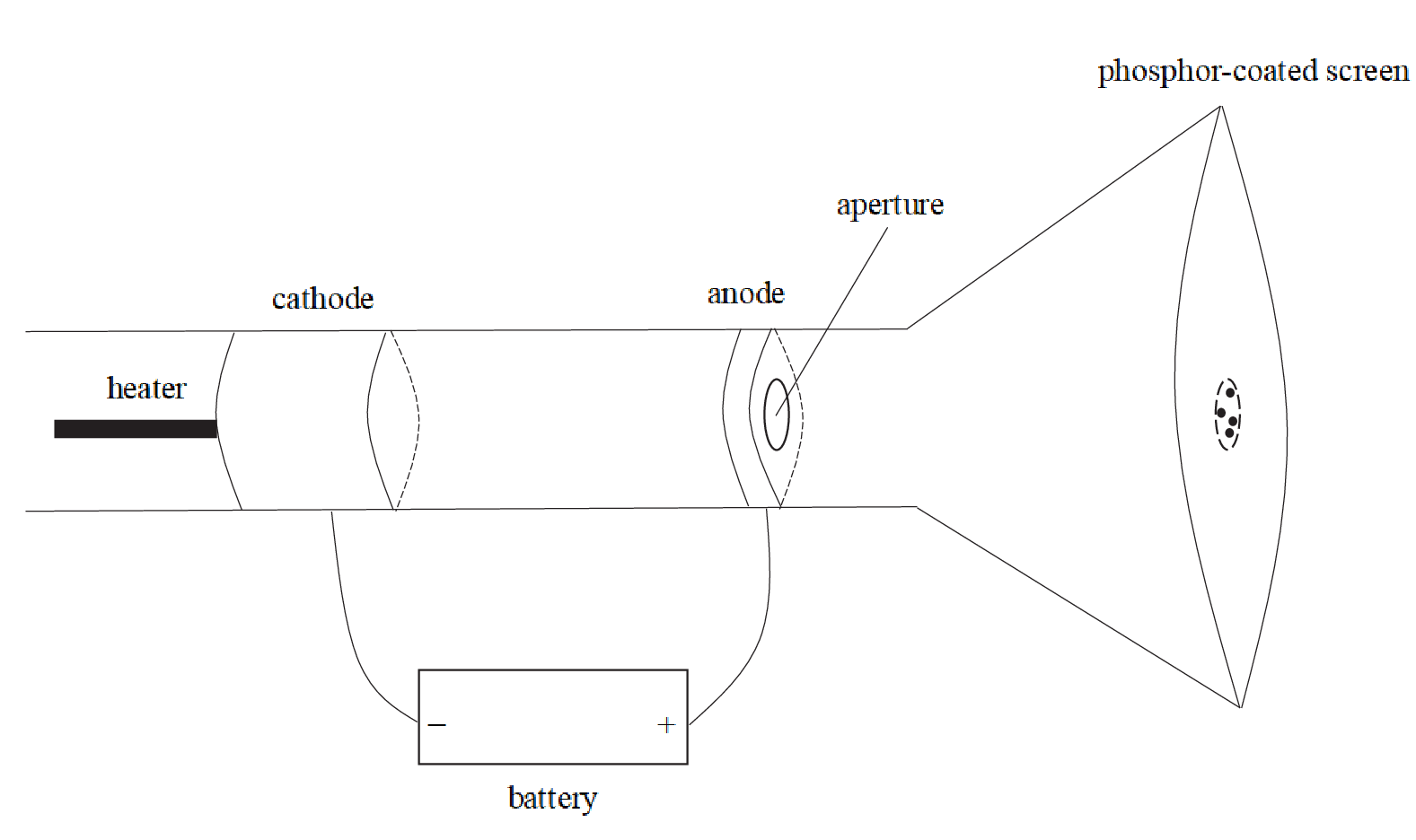}
\caption{Electron gun. The heater warms up the cathode (connected to the negative end of a battery) and electrons are emitted. An aperture is made in the anode (connected to the positive end of the battery). The electron beam passes through the aperture towards the screen. If we turn down the heater, individual flashes appear one at a time with a great interval of time between them.}
\label{fig11}
\end{figure}

\textit{Experiment 2}: We place a barrier after the anode with a single narrow slit (see FIG.12).

\begin{figure}[tb]
\includegraphics[width= 1.0\columnwidth]{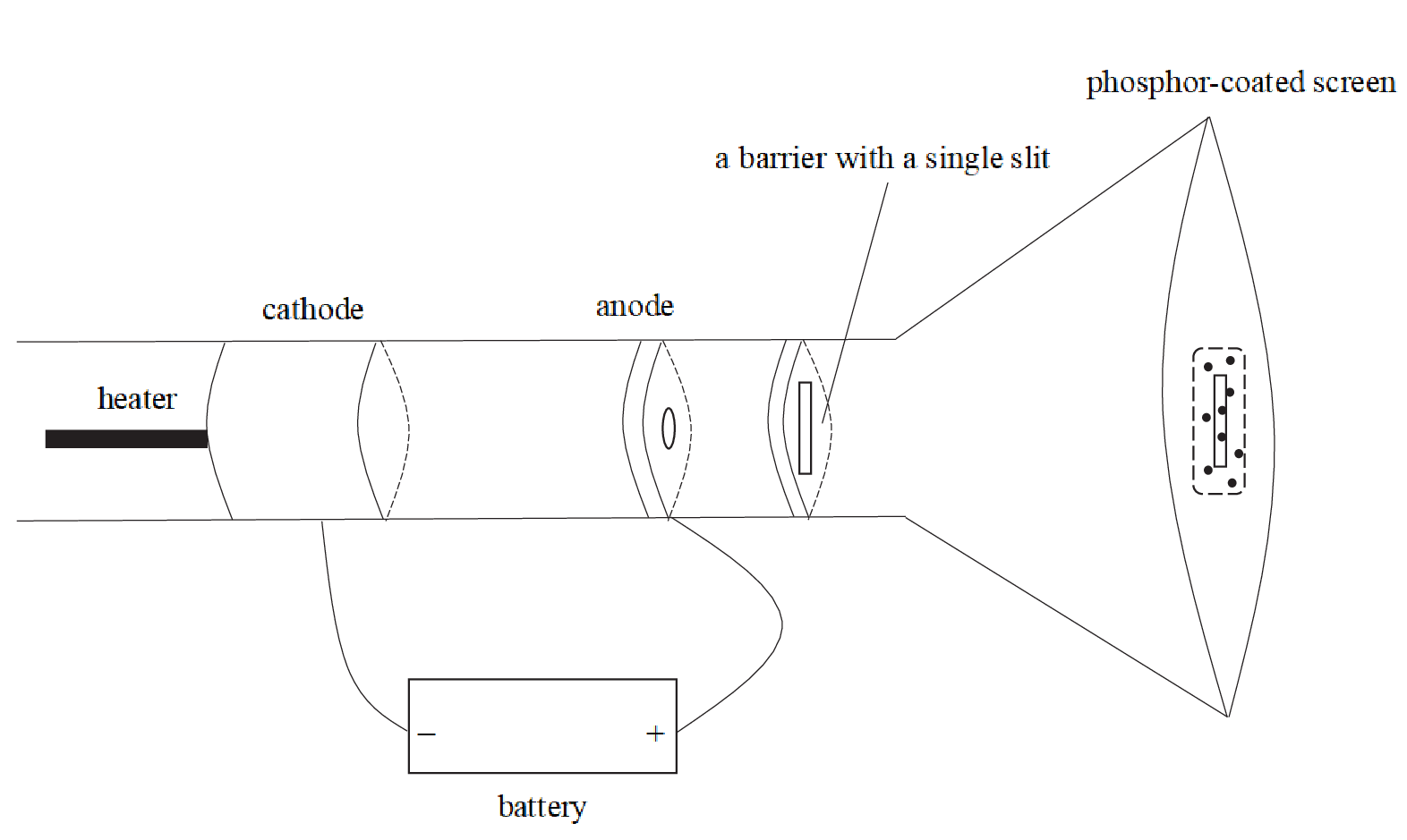}
\caption{Diffraction of single electrons. We place a barrier after the anode with a thin single slit. A bright spot initially is just as narrow as the shape of the slit. But making the slit even thinner, the thin spot on the screen gets wider and spread out. This phenomenon is the familiar diffraction. However, even when we turn down the heater, such that only individual flashes appear, they nonetheless appear on exactly the same place, where the \textit{diffracted} and \textit{widened} spot was.}
\label{fig12}
\end{figure}

We see that the bright spot initially is just as narrow as the shape of the slit. But making the slit even thinner, the thin spot on the screen gets wider and spread out. This phenomenon is the familiar diffraction. However, even when we turn down the heater, such that only individual flashes appear, they nonetheless appear on exactly the same place, where the \textit{diffracted} and \textit{widened} spot was (FIG. 12). This shows that there is a diffraction even with individual flashes. This is unexpected. Indeed, diffraction makes sense for waves, but not for individual flashes.

\textit{Experiment 3}: We place a barrier after the anode with \textit{two} slits (the famous double slit experiment), as shown in FIG.13.

\begin{figure}[tb]
\includegraphics[width= 1.0\columnwidth]{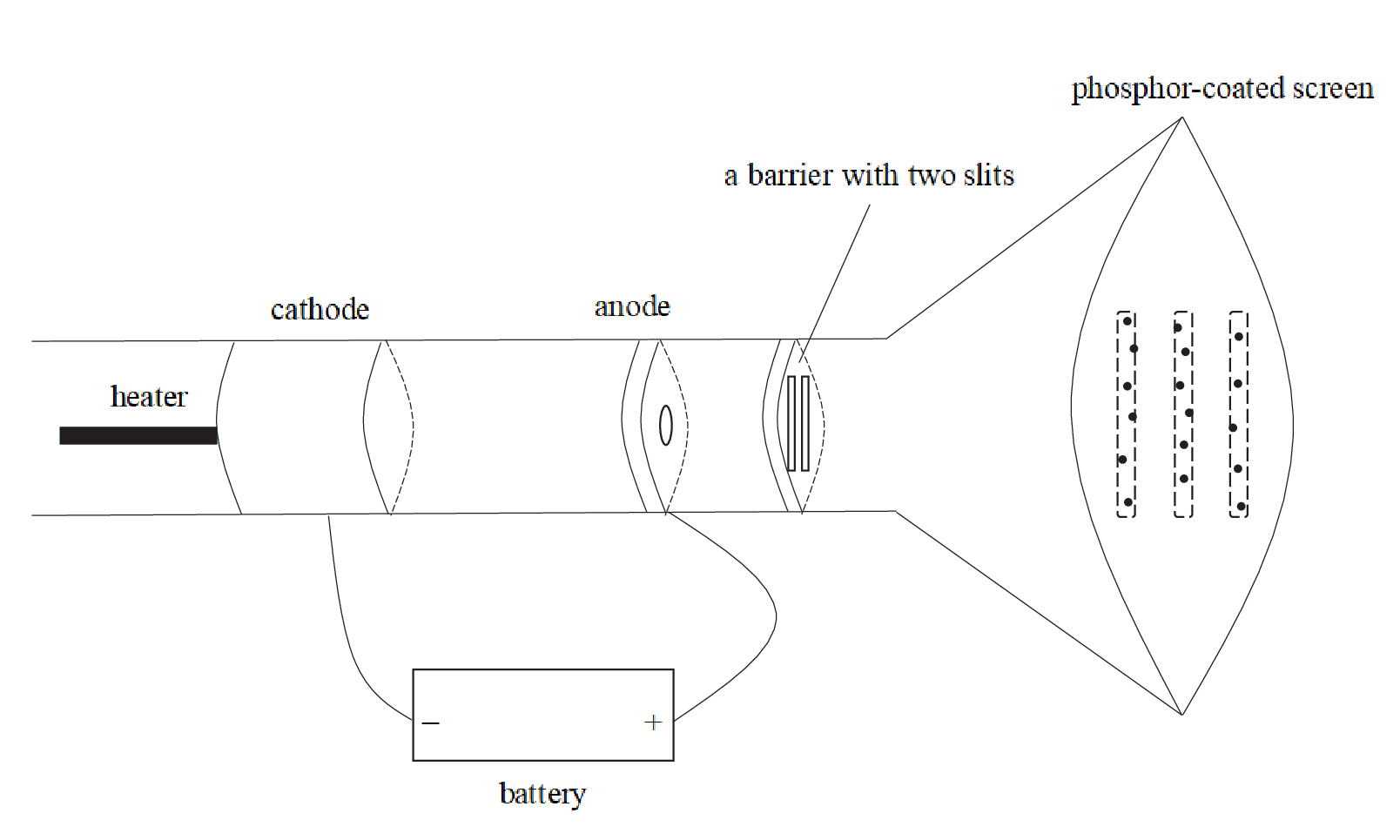}
\caption{Interference of single electrons. We place a barrier with \textit{two} slits after the anode. For sufficiently thin slits, the bright spot on the screen has interference bands - regions of extremely high intensity alternating with quiescent regions. However, even in the case, when we turn down the heating element, such that only individual flashes appear, they nonetheless appear only where the interference maximums are, and never at the location of the interference minimums.}
\label{fig13}
\end{figure}

 We see that for sufficiently thin slits, the bright spot on the screen has interference bands - regions of extremely high intensity alternating with quiescent regions. So far, this is not striking. Interference phenomena are quite familiar. However, the striking thing is that even in the case, when we turn down the heating element, such that only individual flashes appear, they nonetheless appear only where the interference maximums are, and never at the location of the interference minimums (FIG. 13). This is striking, since interference makes sense for waves, but not for individual flashes.

We shall try to explain these three experiments by using modified ideas of de Broglie (modified by Bohm \cite{Bohm1952, Bohm1995, Norsen2017}). He thought that the particles are guided by waves. This is the famous pilot wave theory. In other words, de Broglie thought that there are \textit{both} particles \textit{and} waves. For each particle, there is attached a particular wave, that somehow guides the particle. Particles are never guided to the interference minimum of the wave and more frequently to the interference maximums. Later people removed the particles and their trajectories and only the wave was left. The individual flashes were explained with a sudden collapse of the wave. This was done for reasons of parsimony - in the new theory there are fewerg structures - just a wave and no particles. This theory with only a wave is called Copenhagen interpretation. To this day, pilot wave theory is the best rival of the Copenhagen interpretation \cite{Norsen2017}. However, we do not wish to enter into a dispute about interpretations. We shall simply follow history. For the time being we shall adopt de Broglie's pilot wave theory (perfected by Bohm) and later we shall remove the particles and their trajectories, following Copenhagen interpretation (since we follow the historical development).

The question is this: How can we attach a wave that guides a single particle, such that the three experiments can be explained? We shall make a short and quick review of classical mechanics and of the principle of least action. Later, we shall be able to attach a wave to each particle by comparing the equation for the action $S$ of the particle with the Eikonal equation. This comparison will help us to derive the wave equation for these waves, called Schr\"{o}dinger's equation.

\section{The Principle of Least Action}

Let us examine a single particle moving in 1D (later we shall generalize to 3D motions). It is well known that as the particle moves, it selects that law of motion, for which the action $S$ is stationary \cite{Sommerfeld1964}. The action is given by:
\begin{equation}
S=\int_{t_{1}}^{t_{2}}L(x,\dot{x},t)dt
\end{equation}
where $L=T-U$ is the Lagrangian of the particle. Here $T=m\dot{x}^{2}/2$ is the kinetic energy of the particle, while $U(x)$ is the potential energy. It is well known that the Lagrangian must obey the famous Euler-Lagrange equations \cite{Sommerfeld1964} for $S$ to be stationary:
\begin{equation}
\frac{d}{dt}\left(\frac{\partial L}{\partial \dot{x}}\right)\Bigg|_{x=\overline{x}}=\frac{\partial L}{\partial x}\Bigg|_{x=\overline{x}}\label{Euler-L}
\end{equation}
The overline in $\overline{x}(t)$ designates the \textit{true} law of motion unlike all other \textit{virtual} motions $x(t)$. This equation holds true only when $x=\overline{x}(t)$. In order to derive this equation, people usually consider small virtual deviations from the true trajectory $\overline{x}(t)$ \cite{Sommerfeld1964}. In that case there is only one true trajectory, and a lot of virtual trajectories. 

Let us now consider a \textit{different} scenario. Instead of one real trajectory and many virtual trajectories, we consider \textit{only} \textit{real} trajectories. For instance, we depict two real but close trajectories $\overline{x}(t)$ and $\overline{x}_{1}(t)=\overline{x}(t)+\delta \overline{x}(t)$ in FIG. 14.

\begin{figure}[tb]
\includegraphics[width= 1.0\columnwidth]{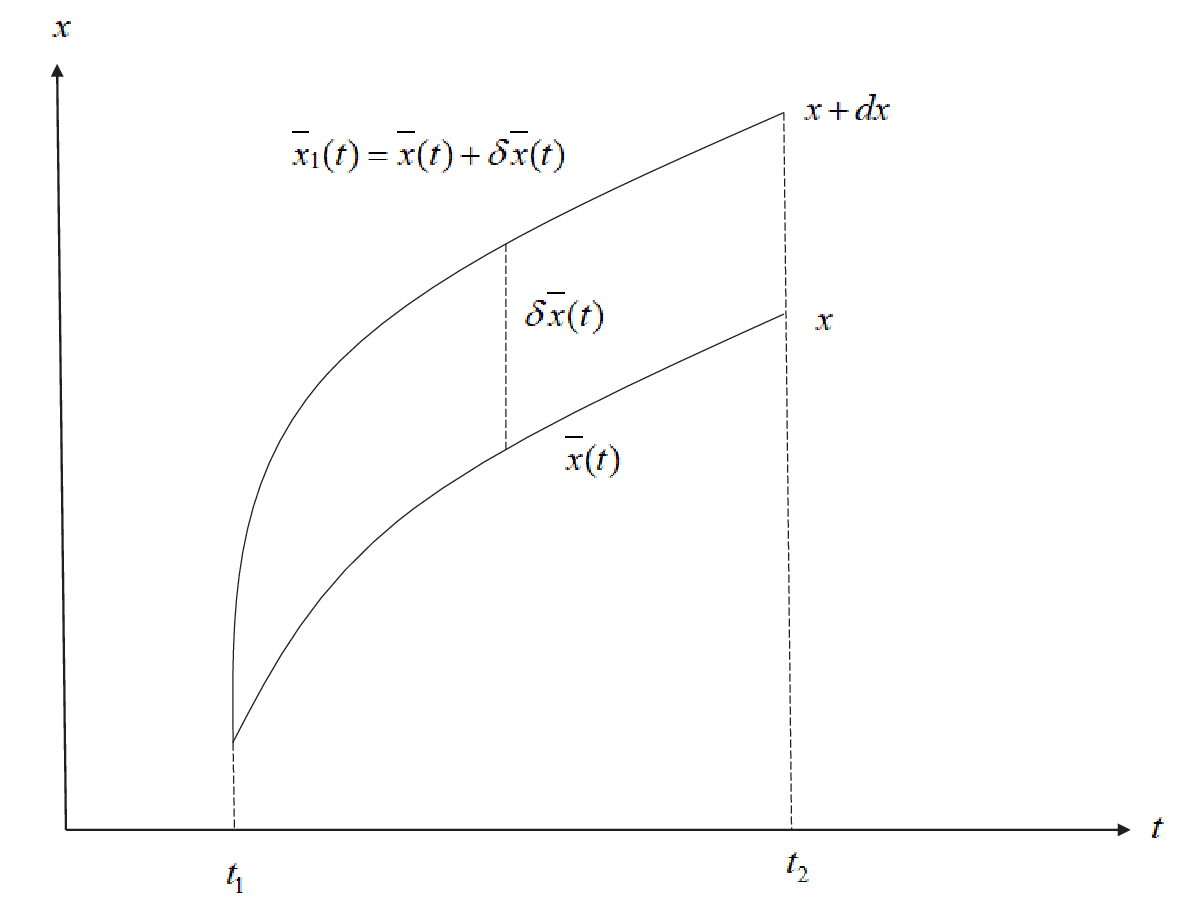}
\caption{The action $S$ along two real trajectories. $S$ will be different at points $x$ and $x+dx$, since it is calculated along different (though real) trajectories $\overline{x}$ and $\overline{x}_{1}$. In order words, the action $S$ will become a function of $x$.}
\label{fig14}
\end{figure}

At the final moment $t=t_{2}$, $\overline{x}(t_{2})=x$ and $\overline{x}_{1}(t_{2})=x+dx$. The action $S$ will be different at points $x$ and $x+dx$, since it is calculated along different (though real) trajectories $\overline{x}$ and $\overline{x}_{1}$. In order words, the action $S$ will become a function of $x$. In that case we have:
\begin{equation}
dS=\frac{\partial S}{\partial x}dx\label{act00}
\end{equation}
On the other hand we have:
\begin{align}
dS&=\int_{t_{1}}^{t_{2}}L(\overline{x}_{1},\dot{\overline{x}}_{1},t)dt-\int_{t_{1}}^{t_{2}}L(\overline{x},\dot{\overline{x}},t)dt\notag\\
&=\int_{t_{1}}^{t_{2}}L(\overline{x}+\delta\overline{x},\dot{\overline{x}}+\delta\dot{\overline{x}},t)dt-\int_{t_{1}}^{t_{2}}L(\overline{x},\dot{\overline{x}},t)dt\notag\\
&=\int_{t_{1}}^{t_{2}}\left[\frac{\partial L}{\partial x}\Bigg|_{x=\overline{x}}\delta\overline{x}+\frac{\partial L}{\partial \dot{x}}\Bigg|_{x=\overline{x}}\delta\dot{\overline{x}}\right]dt\label{act0}
\end{align}
We have Taylor expanded $L(\overline{x}+\delta\overline{x},\dot{\overline{x}}+\delta\dot{\overline{x}},t)$. We also take into account that
\begin{equation}
\delta\dot{\overline{x}}\equiv\dot{\overline{x}}_{1}-\dot{\overline{x}}=\frac{d}{dt}\left[\overline{x}+
\delta\overline{x}\right]-\overline{x}=\frac{d}{dt}\delta\overline{x}.
\end{equation}
We integrate by parts the second term in Eq. \eqref{act0}:
\begin{equation}
\int_{t_{1}}^{t_{2}}\frac{\partial L}{\partial \dot{x}}\Bigg|_{x=\overline{x}}\frac{d}{dt}\delta\overline{x}dt=\frac{\partial L}{\partial \dot{x}}\Bigg|_{x=\overline{x}(t_{2})}\delta\overline{x}(t_{2})-\int_{t_{1}}^{t_{2}}\frac{d}{dt}\frac{\partial L}{\partial \dot{x}}\Bigg|_{x=\overline{x}}dt\label{act1}
\end{equation}
Now, we know that $\delta\overline{x}(t_{2})=dx$. It well known \cite{Sommerfeld1964} that the momentum of the particle $p$ at the point $x$ is:
\begin{equation}
p=\frac{\partial L}{\partial \dot{x}}\Bigg|_{x=\overline{x}(t_{2})}
\end{equation}
In that case Eq. \eqref{act1} becomes:
\begin{equation}
\int_{t_{1}}^{t_{2}}\frac{\partial L}{\partial \dot{x}}\Bigg|_{x=\overline{x}}\frac{d}{dt}\delta\overline{x}dt=pdx-\int_{t_{1}}^{t_{2}}\frac{d}{dt}\frac{\partial L}{\partial \dot{x}}\Bigg|_{x=\overline{x}}\delta\overline{x}(t)dt\label{act2}
\end{equation}
We substitute Eq. \eqref{act2} into Eq. \eqref{act0} and we obtain:
\begin{equation}
dS=pdx+\int_{t_{1}}^{t_{2}}\left[\frac{\partial L}{\partial x}\Bigg|_{x=\overline{x}}-\frac{d}{dt}\frac{\partial L}{\partial \dot{x}}\Bigg|_{x=\overline{x}}\right]\delta\overline{x}(t)dt
\end{equation}
The last integral however is $0$, since for the true motion $x=\overline{x}$ the Lagrangian obeys Euler-Lagrange equations \eqref{Euler-L}. Then we finally have:
\begin{equation}
dS=pdx
\end{equation}
By comparing with Eq.  \eqref{act00} we finally have:
\begin{equation}
p=\frac{\partial S}{\partial x}
\end{equation}
If the motion were 3D, it is easy to show in the same way that: $p_{x}=\frac{\partial S}{\partial x}$, $p_{y}=\frac{\partial S}{\partial y}$ and $p_{z}=\frac{\partial S}{\partial z}$. In vector form, we have:
\begin{equation}
\textbf{p}=\nabla S
\end{equation}
In other words, if we can calculate the function $S(x,y,z)$ we can always calculate the momentum $\textbf{p}$ of the particle, if it passes through the point $\textbf{r}=(x,y,z)$. It is normal to the level-surface $S$, as shown in FIG. 15.

\begin{figure}[tb]
\includegraphics[width= 1.0\columnwidth]{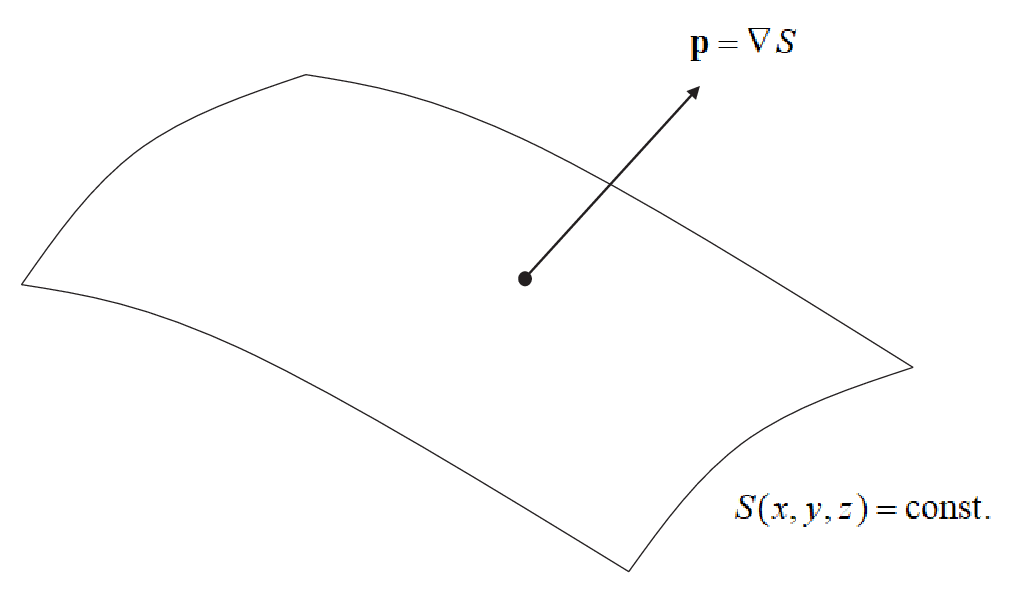}
\caption{The momentum $\textbf{p}=\nabla S$ is perpendicular to the surfaces $S(\textbf{r})=\text{const}.$}
\label{fig15}
\end{figure}
In order to derive the equation for $S$, we shall use the law of conservation of energy: $\textbf{p}^{2}/2m+U(\textbf{r})=E$. We rewrite it in the form:
\begin{equation}
\textbf{p}^{2}=2m\left(E-U(\textbf{r})\right)
\end{equation}
We then substitute $\textbf{p}=\nabla S$ into this equation and we obtain:
\begin{equation}
\left(\nabla S\right)^{2}=2m\left(E-U(\textbf{r})\right)\label{H-J1}
\end{equation}
This is Hamilton-Jacobi equation for a beam of particles with the same energy $E$. We could in principle construct many such surfaces and from them we can construct the trajectory of any particular particle, using the condition $\textbf{p}=\nabla S$. 

However, we can generalize this result. We may consider real trajectories even for different moments in time $t$ and $t+dt$ not just different points $x$ and $x+dx$ (see FIG. 16). 

\begin{figure}[tb]
\includegraphics[width= 1.0\columnwidth]{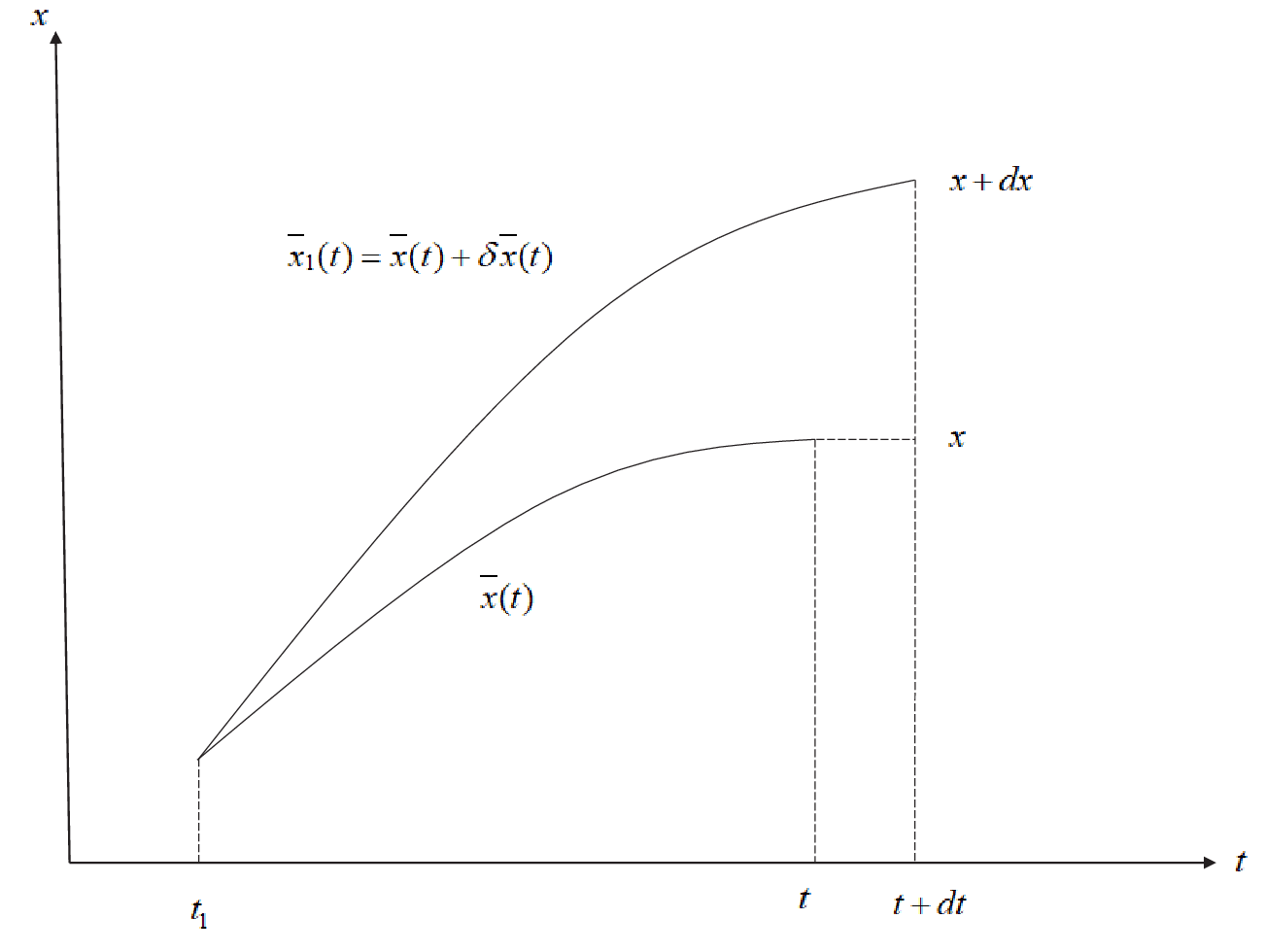}
\caption{The action $S$ as a function of both space and time, i.e. $S=S(x,t)$. $S(x,t)$ is the action that corresponds to the trajectory $\overline{x}(t)$, while the action $S(x+dx,t+dt)$ is the action that corresponds to the \textit{other} \textit{real} trajectory $\overline{x}_{1}(t)$. Here the real trajectory $\overline{x}_{1}(t)=\overline{x}(t)+\delta \overline{x}(t)$ is close to the other real trajectory $\overline{x}(t)$. In this case then the action $S(x,t)$ is function of \textit{both} space \textit{and} time.}
\label{fig16}
\end{figure}

The action $S(x,t)$ is the action that corresponds to the trajectory $\overline{x}(t)$, while the action $S(x+dx,t+dt)$ is the action that corresponds to the \textit{other} \textit{real} trajectory $\overline{x}_{1}(t)$. Here the real trajectory $\overline{x}_{1}(t)=\overline{x}(t)+\delta \overline{x}(t)$ is close to the other real trajectory $\overline{x}(t)$. Since in this case the action $S(x,t)$ is function of \textit{both} space \textit{and} time we have:
\begin{equation}
dS=\frac{\partial S}{\partial x}dx+\frac{\partial S}{\partial t}dt
\end{equation}
On the other hand,
\begin{equation}
dS=\int_{t_{1}}^{t+dt}L\left(\overline{x}_{1},\dot{\overline{x}}_{1},\tau\right)d\tau-\int_{t_{1}}^{t}L\left(\overline{x},\dot{\overline{x}},\tau\right)d\tau\label{leastact0}
\end{equation}
The first integral is:
\begin{equation}
\int_{t_{1}}^{t+dt}L\left(\overline{x}_{1},\dot{\overline{x}}_{1},\tau\right)d\tau=\int_{t_{1}}^{t}L\left(\overline{x}_{1},\dot{\overline{x}}_{1},\tau\right)d\tau+L\left(\overline{x}_{1},\dot{\overline{x}}_{1},t\right)dt\label{leastact1}
\end{equation}
We substitute \eqref{leastact1} into \eqref{leastact0} and we have:
\begin{equation}
dS=\int_{t_{1}}^{t}\left[\ L\left(\overline{x}_{1},\dot{\overline{x}}_{1},\tau\right)-L\left(\overline{x},\dot{\overline{x}},\tau\right)\right] d\tau+L\left(\overline{x}_{1},\dot{\overline{x}}_{1},t\right)dt\label{la00}
\end{equation}
The integral in this equation can be integrated by parts in the same manner as we did the integral in Eq. \eqref{act0}. Then in analogous way we derive:
\begin{align}
&\int_{t_{1}}^{t}\left[\ L\left(\overline{x}_{1},\dot{\overline{x}}_{1},\tau\right)-L\left(\overline{x},\dot{\overline{x}},\tau\right)\right] d\tau=\notag\\
&\frac{\partial L}{\partial \dot{x}}\Bigg|_{x=\overline{x}}\delta\overline{x}(t)+\int_{t_{1}}^{t_{2}}\left[\frac{\partial L}{\partial x}\Bigg|_{x=\overline{x}}-\frac{d}{d\tau}\frac{\partial L}{\partial \dot{x}}\Bigg|_{x=\overline{x}}\right]\delta\overline{x}(\tau)d\tau\label{la11}
\end{align}
But again, $\partial L/\partial\dot{x}$ is the momentum $p$ of the particle, while the integral in this equation is $0$ (the Lagrangian for the true law of motion obeys Euler-Lagrange equations). Then, if we substitute \eqref{la11} into \eqref{la00} we derive:
\begin{equation}
dS=p\delta\overline{x}(t)+L\left(\overline{x}_{1}(t),\dot{\overline{x}}_{1}(t),t\right)dt\label{leastact2}
\end{equation}
We transform the last term:
\begin{align}
&L\left(\overline{x}_{1}(t),\dot{\overline{x}}_{1}(t),t\right)dt=L\left(\overline{x}(t)+\delta \overline{x},\dot{\overline{x}}(t)+\delta\dot{\overline{x}},t\right)dt\notag\\
&=L\left(\overline{x}(t),\dot{\overline{x}}(t),t\right)dt+\frac{\partial L}{\partial x}\Bigg|_{x=\overline{x}}\delta\overline{x}dt+\frac{\partial L}{\partial \dot{x}}\Bigg|_{x=\overline{x}}\delta\dot{\overline{x}}dt
\end{align}
However, the last two expressions are second order and we can ignore them. Then we have that:
\begin{equation}
L\left(\overline{x}_{1}(t),\dot{\overline{x}}_{1}(t),t\right)dt=L\left(\overline{x}(t),\dot{\overline{x}}(t),t\right)dt
\end{equation}
When we substitute this result into Eq. \eqref{leastact2} we obtain:
\begin{equation}
dS=p\delta\overline{x}(t)+L\left(\overline{x}(t),\dot{\overline{x}}(t),t\right)dt\label{leastact3}
\end{equation}
We wish to connect $\delta\overline{x}(t)$ with $dx$. Note that (FIG. 16) $\delta\overline{x}(t)\neq dx$ but it is easy to see that:
\begin{equation}
dx=\delta\overline{x}+\dot{\overline{x}}_{1}dt\approx \delta\overline{x}+\dot{\overline{x}}dt
\end{equation}
In that case we have:
\begin{equation}
\delta\overline{x}=dx-\dot{\overline{x}}dt
\end{equation}
We substitute this result into Eq. \eqref{leastact3} and we finally have:
\begin{equation}
dS=pdx-\left(p\dot{\overline{x}}-L\right)dt
\end{equation}
It is well known \cite{Sommerfeld1964} that the term in the brackets is the energy $E$ of the particle. Then we finally have:
\begin{equation}
dS=pdx-Edt
\end{equation}
If the motion were 3D, in an analogous manner we would have obtained:
\begin{equation}
dS=p_{x}dx+p_{y}dy+p_{z}dz-Edt
\end{equation}
In other words,
\begin{equation}
\frac{\partial S}{\partial x}dx+\frac{\partial S}{\partial y}dy+\frac{\partial S}{\partial z}dz+\frac{\partial S}{\partial t}dt=p_{x}dx+p_{y}dy+p_{z}dz-Edt
\end{equation}
This leads us to conclude that:
\begin{equation}
\textbf{p}=\nabla S
\end{equation}
as well as:
\begin{equation}
E=-\frac{\partial S}{\partial t}
\end{equation}
If we plug these results into the energy conservation law $E=\textbf{p}^{2}/2m+U$, we have:
\begin{equation}
-\frac{\partial S}{\partial t}=\frac{1}{2m}(\nabla S)^{2}+U(\textbf{r})\label{H-J2}
\end{equation}
This is the famous Hamilton-Jacobi equation \cite{Sommerfeld1964}.

\section{Comparison between the Eikonal equation and the Hamilton-Jacobi equation. Derivation of Schr\"{o}dinger's equation}

\subsection{Comparison between the Eikonal equation and the Hamilton-Jacobi equation}

The attentive reader may have noticed quite a great resemblance between the Eikonal equations \eqref{eikonal} and \eqref{eikonal2} on the one hand and the Hamilton-Jacobi's equations \eqref{H-J1} and \eqref{H-J2} on the other. We can also see this comparison in Table I.

\begin{table}[h]
\caption{Comparison of the Eikonal equations in optics and the Hamilton-Jacobi equations in mechanics} 

\begin{tabular}{|l|c|c|}\hline
 & optics & mechanics \\ \hline
 & & \\
Wave Equation & $\nabla^{2}f-\frac{n^{2}}{c^{2}}\frac{\partial ^{2}f}{\partial t^{2}}=0$ & ? \\
& &\\
& $f=\text{Re}\left[A(\textbf{r})e^{iS(\textbf{r})}\right]$& \\ 
& & \\
\hline
& & \\
regime of $\lambda\rightarrow 0$ & $\left(\nabla S\right)^{2}=\frac{n^{2}\omega^{2}}{c^{2}}$ & $\left(\nabla S\right)^{2}=2m(E-U)$ \\
no $t$ dependence & &\\ \hline
 & & \\
regime of $\lambda\rightarrow 0$&  $\left(\nabla S\right)^{2}=\frac{n^{2}}{c^{2}}\left(\frac{\partial S}{\partial t}\right)^{2}$ & $\left(\nabla S\right)^{2}=-2m\left(\frac{\partial S}{\partial t}+U\right)$ \\
$t$ dependence& & \\ \hline
\end{tabular}
\end{table}

Now, when we examine Table I and consider the resemblance between the Eikonal equations of geometrical optics and Hamilton-Jacobi equations in Newtonian mechanics, it is quite natural to ask the following question: is it the case that the Hamilton-Jacobi equations (which reproduce the familiar Newtonian mechanics) are just the regime of $\lambda\rightarrow 0$ of some yet unknown wave equation? In other words, could it be the case that the Hamilton-Jacobi's equations \textit{are} the Eikonal equations of some unknown wave equation? And can we find out what this wave equation is? Perhaps this wave equation can explain somehow the electron gun experiments with interference and diffraction of single particles!

When we compare the Eikonal equations with the Hamilton-Jacobi equations (Table I), we see that the action $S$ resembles a phase of unknown wave. Then, if we designate the unknown wave with $\Psi(\textbf{r},t)$, it is tempting to postulate that: $\Psi(\textbf{r},t)=Ae^{iS}$, where $S(\textbf{r},t)$ is the action in classical mechanics. However the action $S$ has units of energy$\times$time, while the phase of any wave should be dimensionless. Therefore, the unknown wave ought to be of the form:
\begin{equation}
\Psi(\textbf{r},t)=A(\textbf{r},t)e^{iS/\hbar}
\end{equation}
Here $\hbar$ is some quantity that should have a dimension of action. Only then $S/\hbar$ is dimensionless. It will turn out (see below) that $\hbar$ is the familiar Planck's constant. We also expect that the unknown wave equation for $\Psi$ will be reduced to the Hamilton-Jacobi equations when $S/\hbar\gg 1$. Indeed, we have shown in Eq. \eqref{largephase} that for large phases of the waves (in our case $S/\hbar\gg 1$), the wave equation is automatically reduced to the regime of $\lambda\rightarrow 0$. In other words, we shall treat the condition $S/\hbar\gg 1$ as equivalent to $\lambda\rightarrow 0$.

But even in the regime $\lambda\rightarrow 0$ ($S/\hbar\gg 1$) the yet unknown wave $\Psi(\textbf{r},t)$ will \textit{not} give individual flashes. Indeed, the regime $\lambda\rightarrow 0$ ($S/\hbar\gg 1$) simply says that a parallel beam (which may be very thick) of the wave $\Psi(\textbf{r},t)$ moves in a rectilinear way (as in the case of geometrical optics). If we are to have \textit{individual} flashes we need more. In order to explain the individual flashes, De Broglie thought that there are \textit{both} particles \textit{and} waves. The individual flashes \textit{are} the particles, which are guided by the phase of the wave in this manner: if we somehow know the wave $\Psi=Ae^{iS/\hbar}$ and its phase $S/\hbar$, we can calculate the momentum $\textbf{p}$ of the particle by the guiding equation $\textbf{p}=\nabla S$.

\textit{Summary}: De Broglie's idea to explain the electron gun experiments with interference and diffraction is this: To each particle, attach a wave $\Psi(\textbf{r},t)=Ae^{iS/\hbar}$ and guide the particle by the guiding equation: $\textbf{p}=\nabla S$. The wave will obey some yet unknown wave equation. In the limit $\lambda\rightarrow 0$ ($S/\hbar\gg 1$), the unknown wave equation will be reduced to its Eikonal equations, which are in fact the Hamilton-Jacobi equations shown in Table I. This regime of $\lambda\rightarrow 0$ ($S/\hbar\gg 1$) then will give us the familiar Newtonian mechanics. However, we hope that if we are \textit{not} in the regime $\lambda\rightarrow 0$ ($S/\hbar\gg 1$), the wave $\Psi$ will guide the particles in such a way, that the electron gun experiments with interference and diffraction \textit{can} be explained. What is left then is to find the unknown wave equation for $\Psi$.

Before we do that, let us transform the guiding equation $\textbf{p}=\nabla S$. Let us take $\nabla \Psi/\Psi$:
\begin{equation}
\frac{\nabla\Psi}{\Psi}=\frac{(\nabla A+\frac{i}{\hbar}A\nabla S)e^{iS/\hbar}}{Ae^{iS/\hbar}}=\frac{\nabla A}{A}+\frac{i}{\hbar}\nabla S
\end{equation}
Then $\nabla S$ is given by:
\begin{equation}
\nabla S =\hbar \;\text{Im}\left(\frac{\nabla\Psi}{\Psi}\right)=\frac{\hbar}{2i}\frac{\Psi^{*}\nabla\Psi-\Psi\nabla\Psi^{*}}{\Psi^{*}\Psi}
\end{equation}
We have used the fact that for any complex number $z$, $\text{Im}z=(z-z^{*})/2i$. The particle's velocity is then given by:
\begin{equation}
\textbf{v}=\frac{1}{m}\nabla S=\frac{\hbar}{2mi}\frac{\Psi^{*}\nabla\Psi-\Psi\nabla\Psi^{*}}{\Psi^{*}\Psi}
\end{equation}

\subsection{Derivation of the wave equation from the Eikonal equation}

Thus far we were able to derive the \textit{Eikonal} equations of geometrical optics \textit{from} the \textit{wave} equation of wave optics (see Table I). Now we wish to do the \textit{opposite} procedure: from the Eikonal equations of geometrical optics we shall guess what is the wave equation of wave optics.  This will be important for us later when we derive Schr\"{o}dinger's wave equation for the wave $\Psi$.

Let us consider a complex wave $F$ of the type:
\begin{equation}
F=A(\textbf{r},t)e^{iS(\textbf{r},t)}
\end{equation}
We pretend that we do not know the wave equation for $F$, Eq. \eqref{waveeqn}. We only know that the phase $S$ obeys the time-dependent Eikonal equation \eqref{eikonal2}:
\begin{equation}
\left(\nabla S\right)^{2}=\frac{n^{2}}{c^{2}}\left(\frac{\partial S}{\partial t}\right)^{2}\label{eikonal123}
\end{equation}
This equation is the regime of geometrical optics of the unknown wave equation. This regime means that the phase $S$ is quite large and is of the type:
\begin{equation}
S=\tilde{S}/\epsilon
\end{equation}
where $\tilde{S}\sim 1$, while $\epsilon \ll 1$. Then $S=\tilde{S}/\epsilon\gg 1$.

Let us take the following derivative:
\begin{equation}
\frac{\partial}{\partial x}\left[Ae^{iS}\right]=\left(\frac{\partial A}{\partial x}+iA\frac{\partial S}{\partial x}\right)e^{iS}
\end{equation}
In the limit of geometrical optics, the last term is of the order of $\epsilon^{-1}$ and is thus the largest. In that case we can ignore the first term and we get:
\begin{equation}
\frac{\partial}{\partial x}\left[Ae^{iS}\right]\approx iA\frac{\partial S}{\partial x}e^{iS}
\end{equation}
In other words, we can replace:
\begin{equation}
\frac{\partial S}{\partial x}Ae^{iS}\approx \frac{1}{i}\frac{\partial}{\partial x}\left[Ae^{iS}\right]
\end{equation}
In the same way it is very easy to prove that for the square of the first derivative we have:
\begin{equation}
\left(\frac{\partial S}{\partial x}\right)^{2} Ae^{iS}\approx\left(\frac{1}{i}\frac{\partial}{\partial x}\right)^{2}\left[Ae^{iS}\right]\label{go-we1}
\end{equation}
Indeed, we have
\begin{equation}
\frac{\partial ^{2}}{\partial x^{2}}\left[Ae^{iS}\right]=\left[\frac{\partial ^{2}A}{\partial x^{2}}+2i\frac{\partial A}{\partial x}\frac{\partial S}{\partial x}+iA\frac{\partial ^{2}S}{\partial x^{2}}-A\left(\frac{\partial S}{\partial x}\right)^{2}\right]e^{iS}
\end{equation}
In the geometrical optics regime, the last term is the largest, since it is of the order of $\epsilon^{-2}$. Then, we can ignore all other terms and we get:
\begin{equation}
\frac{\partial ^{2}}{\partial x^{2}}\left[Ae^{iS}\right]\approx -A\left(\frac{\partial S}{\partial x}\right)^{2}e^{iS}
\end{equation}
which proves Eq. \eqref{go-we1}. Eq. \eqref{go-we1} is also valid for the other derivatives $y$, $z$ and the time $t$. With this preparation at hand, let us start with the time-dependent Eikonal equation in geometrical optics \eqref{eikonal123}, written with partial derivatives:
\begin{equation}
\left(\frac{\partial S}{\partial x}\right)^{2}+\left(\frac{\partial S}{\partial y}\right)^{2}+\left(\frac{\partial S}{\partial z}\right)^{2}=\frac{n^{2}}{c^{2}}\left(\frac{\partial S}{\partial t}\right)^{2}
\end{equation}
We recall that we pretend we do not know the wave equation \eqref{waveeqn}. We wish to guess it from the above equation. To this end, let us multiply the above equation with $Ae^{iS}$:
\begin{align}
&\left(\frac{\partial S}{\partial x}\right)^{2}Ae^{iS}+\left(\frac{\partial S}{\partial y}\right)^{2}Ae^{iS}+\left(\frac{\partial S}{\partial z}\right)^{2}Ae^{iS}\notag\\
&=\frac{n^{2}}{c^{2}}\left(\frac{\partial S}{\partial t}\right)^{2}Ae^{iS}\label{Eikonal124}
\end{align}
Since, the above equation is the limit of geometrical optics, we can use approximation \eqref{go-we1} and then Eq. \eqref{Eikonal124} is transformed into:
\begin{align}
&\left(\frac{1}{i}\frac{\partial}{\partial x}\right)^{2}\left[Ae^{iS}\right]+\left(\frac{1}{i}\frac{\partial}{\partial y}\right)^{2}\left[Ae^{iS}\right]+\left(\frac{1}{i}\frac{\partial}{\partial z}\right)^{2}\left[Ae^{iS}\right]\notag\\
&=\frac{n^{2}}{c^{2}}\left(\frac{1}{i}\frac{\partial}{\partial t}\right)^{2}\left[Ae^{iS}\right]
\end{align}
If we substitute $F=Ae^{iS}$, the above equation is reduced to:
\begin{equation}
-\nabla^{2}F=-\frac{n^{2}}{c^{2}}\frac{\partial ^{2}}{\partial t^{2}}F
\end{equation}
After some rearrangement, this is precisely the wave equation:
\begin{equation}
\nabla^{2}F-\frac{n^{2}}{c^{2}}\frac{\partial ^{2}F}{\partial t^{2}}=0
\end{equation}
It is amazing, that we were able to guess what is the wave equation from the Eikonal equation. Now we shall follow the same procedure to derive the unknown wave equation for $\Psi$ out of Hamiton-Jacobi equation.

\subsection{Derivation of Schr\"{o}dinger's equaiton}
We consider a wave $\Psi$ of the type:
\begin{equation}
\Psi(\textbf{r},t)=A(\textbf{r},t)e^{\frac{i}{\hbar}S(\textbf{r},t)}
\end{equation}
We do not yet know the wave equation for $\Psi$. But we know that in the regime of 'geometrical optics' (i.e. Newtonian mechanics), i.e. when $S/\hbar\gg 1$ the unknown wave equation is reduced to its 'Eikonal equation', i.e. to Hamilton-Jacobi equation:
\begin{equation}
-\frac{\partial S}{\partial t}=\frac{1}{2m}(\nabla S)^{2}+U(\textbf{r})\label{H-J123}
\end{equation}
The regime of 'geometrical optics' (i.e. Newtonian mechanics) $S/\hbar\gg 1$, means that the Planck's constant $\hbar$ is really small. In this regime of $S/\hbar\gg 1$, we have:
\begin{equation}
\frac{\partial}{\partial x}\left[Ae^{iS/h}\right]=\left(\frac{\partial A}{\partial x}+\frac{i}{\hbar}A\frac{\partial S}{\partial x}\right)e^{iS/\hbar}\approx \frac{i}{\hbar}A\frac{\partial S}{\partial x}e^{iS/\hbar}
\end{equation}
We have left only the largest term of order $\hbar^{-1}$.

In other words, we have the following approximation (valid only when $S/\hbar\gg 1$):
\begin{equation}
\frac{\partial S}{\partial x}Ae^{iS/\hbar}\approx\frac{\hbar}{i}\frac{\partial}{\partial x}\left[Ae^{iS/h}\right]\label{app000}
\end{equation}
In the same way, it is trivial to show that for the \textit{square} of the derivative we have:
\begin{equation}
\left(\frac{\partial S}{\partial x}\right)^{2}Ae^{iS/\hbar}\approx\left(\frac{\hbar}{i}\frac{\partial}{\partial x}\right)^{2}\left[Ae^{iS/h}\right]\label{app111}
\end{equation}
Now, that we have these approximations, we are finally ready to derive Shr\"{o}dinger's equation from the Hamilton-Jacobi equation. Indeed, let us start with the Hamilton-Jacobi equation \eqref{H-J123} in terms of derivatives:
\begin{equation}
-\frac{\partial S}{\partial t}=\frac{1}{2m}\left(\frac{\partial S}{\partial x}\right)^{2}+\frac{1}{2m}\left(\frac{\partial S}{\partial y}\right)^{2}+\frac{1}{2m}\left(\frac{\partial S}{\partial z}\right)^{2}+U
\end{equation}

We multiply both sides with the complex wave $\Psi=Ae^{iS/\hbar}$:
\begin{align}
&-\frac{\partial S}{\partial t}Ae^{iS/\hbar}=\frac{1}{2m}\left(\frac{\partial S}{\partial x}\right)^{2}Ae^{iS/\hbar}+\frac{1}{2m}\left(\frac{\partial S}{\partial y}\right)^{2}Ae^{iS/\hbar}\notag\\
&+\frac{1}{2m}\left(\frac{\partial S}{\partial z}\right)^{2}Ae^{iS/\hbar}+UAe^{iS/\hbar}\label{H-J124}
\end{align}
Next, we use our approximations Eqs. \eqref{app000} and \eqref{app111} but also for $y$, $z$ and $t$ derivatives and Eq. \eqref{H-J124} is transformed into:
\begin{align}
&-\frac{\hbar}{i}\frac{\partial}{\partial t}\left[Ae^{iS/\hbar}\right]=\notag\\
=&\frac{1}{2m}\left(\frac{\hbar}{i}\frac{\partial}{\partial x}\right)^{2}\left[Ae^{iS/h}\right]+\frac{1}{2m}\left(\frac{\hbar}{i}\frac{\partial}{\partial y}\right)^{2}\left[Ae^{iS/h}\right]\notag\\
&+\frac{1}{2m}\left(\frac{\hbar}{i}\frac{\partial}{\partial z}\right)^{2}\left[Ae^{iS/h}\right]+UAe^{iS/\hbar}
\end{align}
After we simplify and we substitute $\Psi=Ae^{iS/\hbar}$ we finally have the famous Schr\"{o}dinger's wave equation:
\begin{equation}
i\hbar\frac{\partial\Psi}{\partial t}=-\frac{\hbar^{2}}{2m}\nabla^{2}\Psi+U\Psi\label{Schr}
\end{equation}
It is this equation that in the regime of `geometrical optics` is reduced to the Hamilton-Jacobi equation \eqref{H-J123}.

\section{Copenhagen interpretation vs Pilot Wave theory}
\subsection{Copenhagen interpretation}
We finally have the complete pilot wave theory of de Broglie-Bohm. We start with the Schr\"{o}dinger's equation for the complex wave $\Psi(\textbf{r},t)=A(\textbf{r},t)e^{iS(\textbf{r},t)/\hbar}$:
\begin{equation}
i\hbar\frac{\partial\Psi}{\partial t}=-\frac{\hbar^{2}}{2m}\nabla^{2}\Psi+U\Psi\label{Schr}
\end{equation}
which guides the particle by the guiding equation:
\begin{equation}
\textbf{v}=\frac{1}{m}\nabla S=\frac{\hbar}{2mi}\frac{\Psi^{*}\nabla\Psi-\Psi\nabla\Psi^{*}}{\Psi^{*}\Psi}\label{guide}
\end{equation}
If we know the initial position of the particle and the initial wave function $\Psi(\textbf{r},t=0)$ we can in principle solve Schr\"{o}dinger's equation \eqref{Schr} and from the guiding condition \eqref{guide} we can determine the trajectory of the particle. We have not yet shown that this theory can explain the electron gun experiments (it can). However, in the limit of geometrical optics $S/\hbar\gg 1$, Schr\"{o}dinger's equation is reduced to the time-dependent Hamilton-Jacobi equation \eqref{H-J123} and given the guiding condition Eq. \eqref{guide} we restore the standard classical Newtonian physics. 

However, for various historical and philosophical reasons \cite{Norsen2017} people decided to explain the individual flashes in the electron gun experiments in a different way. Instead of particles guided by waves, people preferred to explain the individual flashes by a sudden 'collapse' of the wave into a small volume (flash). The probability density for the collapse is postulated to be given by the square of the modulus of the wave function (Born's rule). In other words $|\Psi(\textbf{r},t)|^{2}dV$ gives the probability for the wave to give a flash within a small volume $dV$ centered at a point $\textbf{r}$. In that case the wave function ought to be normalized to unity, i.e. we must require that:
\begin{equation}
\int_{V_{\infty}}|\Psi(\textbf{r},t)|^{2}dV=1\label{norm}
\end{equation}
where $V_{\infty}$ is the whole of space. This ensures that the total probability to find a particle somewhere is $1$. Copenhagen interpretation can now explain the electron gun experiments. Indeed, as the wave function propagates, and passes through a small aperture (see FIG. 12), the wave is diffracted and gives non-zero probability to suddenly collapse in the widened and diffracted spot. This explains Experiment 2. Experiment 3 can also be explained. As the wave passes through the two slits, it creates interference bands so typical for waves. At the minimums, there is a zero probability for a collapse, since there $\Psi=0$ and no flash could appear there. At the interference maximums, there is highest probability for a sudden collapse (flash), since there $|\Psi|^{2}$ is a maximum.

A possible objection to the Copenhagen interpretation is that, if the wave is normalized to unity \eqref{norm} at some moment of time $t_{0}$, we have no guarantee that it will remain normalized to unity for all moments of time $t> t_{0}$. However this objection can be easily answered. Indeed, it is easy to show that if the wave function is normalized to unity at $t_{0}$, it will remain so forever. To this end let us take:
\begin{equation}
\frac{\partial}{\partial t}|\Psi|^{2}=\frac{\partial}{\partial t}\left(\Psi^{*}\Psi\right)=\frac{\partial\Psi^{*}}{\partial t}\Psi+\Psi^{*}\frac{\partial\Psi}{\partial t}\label{Schr0}
\end{equation}

But, from Schr\"{o}dinger's equation \eqref{Schr} we have:
\begin{equation}
\frac{\partial \Psi}{\partial t}=\frac{i\hbar}{2m}\nabla^{2}\Psi-\frac{i}{\hbar}U\Psi\label{Schr1}
\end{equation}
We take complex conjugate of this equation and we have:
\begin{equation}
\frac{\partial \Psi^{*}}{\partial t}=-\frac{i\hbar}{2m}\nabla^{2}\Psi^{*}+\frac{i}{\hbar}U\Psi^{*}\label{Schr2}
\end{equation}
(we assumed that $U$ is real). We substitute Eqs. \eqref{Schr1} and \eqref{Schr2} into Eq. \eqref{Schr0} and we obtain:
\begin{align}
\frac{\partial}{\partial t}|\Psi|^{2}&=-\frac{i\hbar}{2m}\left[\Psi^{*}\nabla^{2}\Psi-\Psi\nabla^{2}\Psi^{*}\right]\notag\\
&=-\nabla\cdot\left[\frac{i\hbar}{2m}\left(\Psi^{*}\nabla\Psi-\Psi\nabla\Psi^{*}\right)\right]
\end{align}
Here we have used that for two arbitrary functions $u(\textbf{r})$ and $v(\textbf{r})$, we have that $u\nabla^{2}v-v\nabla^{2}u=\nabla\cdot(u\nabla v-v\nabla u)$.
If we designate with $\textbf{J}$ (the so called \textit{probability current}):
\begin{equation}
\textbf{J}=\frac{i\hbar}{2m}\left(\Psi^{*}\nabla\Psi-\Psi\nabla\Psi^{*}\right)\label{Schr123}
\end{equation}
we finally obtain that:
\begin{equation}
\frac{\partial}{\partial t}|\Psi|^{2}=-\nabla\cdot \textbf{J}\label{Schr124}
\end{equation}
We are now ready to prove that if the normalization condition \eqref{norm} is true at some arbitrary initial moment of time $t_{0}$, then it will be true for \textit{all} moments of time. Indeed, let us asuume that:
\begin{equation}
\int_{V_{\infty}}|\Psi(\textbf{r},t_{0})|^{2}dV=1
\end{equation}
for $t=t_{0}$. Question: is it true that:
\begin{equation}
\int_{V_{\infty}}|\Psi(\textbf{r},t)|^{2}dV=1\label{norm1}
\end{equation}
for all $t>t_{0}$? To prove that let us take the time derivative of \eqref{norm1}:
\begin{equation}
\frac{d}{dt}\int_{V_{\infty}}|\Psi(\textbf{r},t)|^{2}dV=\int_{V_{\infty}}\frac{\partial}{\partial t}|\Psi(\textbf{r},t)|^{2}=-\int_{V_{\infty}}\nabla\cdot \textbf{J}dV
\end{equation}
In the last line we have used Eq. \eqref{Schr124}. If we use Gauss' theorem:
\begin{equation}
\int_{V_{\infty}}\nabla\cdot \textbf{J}dV=\oint_{S_{\infty}}\textbf{J}\cdot d\textbf{s}=0
\end{equation}
Here the integration $\oint_{S_{\infty}}$ is around the `infinite sphere`, i.e. we integrate through a sphere with some finite radius $R$ and after that we take the limit $R\rightarrow\infty$. This surface integral is $0$, since we assume that the wave function $\Psi$ as well as its derivatives go to $0$ at infinity (square integrability). If that is true, then
\begin{equation}
\frac{d}{dt}\int_{V_{\infty}}|\Psi(\textbf{r},t)|^{2}dV=0
\end{equation}
and this means that:
\begin{equation}
\int_{V_{\infty}}|\Psi(\textbf{r},t)|^{2}dV=\text{const.}
\end{equation}
Then it follows that if $\int_{V_{\infty}}|\Psi(\textbf{r},t_{0})|^{2}dV=1$ at some initial moment of time $t_{0}$, then it will remain true for all moments of time $t>t_{0}$. 

Therefore the Copenhagen interpretation is consistent with $|\Psi|^{2}$ giving probability density. And since this interpretation is more `economical`, i.e. it has only a wave-function and no particles, people have preferred it historically. However, there are still some serious objections to the theory like the ontological problem, the measurement problem and the locality problem \cite{Norsen2017} but we will not enter into such disputes.

\subsection{Pilot wave theory}
For completeness' sake we shall prove in this subsection that the original pilot wave theory can \textit{also} explain the electron gun experiment and it also gives that $|\Psi|^{2}$ is probability density.

\textit{Theorem}: 

Let us have a particle, which is guided by the guiding condition:

\begin{equation}
\textbf{v}=\frac{1}{m}\nabla S=\frac{\hbar}{2mi}\frac{\Psi^{*}\nabla\Psi-\Psi\nabla\Psi^{*}}{\Psi^{*}\Psi}\label{guide000}
\end{equation}
where $\Psi=Ae^{iS/\hbar}$ is a solution of the Schr\"{o}dinger's equation \eqref{Schr}. Then if the probability density $\rho(\textbf{r},t_{0})$ for the \textit{initial} position of the particle at the initial moment $t_{0}$ is given by
\begin{equation}
\rho(\textbf{r},t_{0})=|\Psi(\textbf{r},t_{0})|^{2}
\end{equation}
then for \textit{all} moments of time it remains true that:
\begin{equation}
\rho(\textbf{r},t)=|\Psi(\textbf{r},t)|^{2}\;\;\;\text{for}\; t>t_{0}
\end{equation}

\textit{Proof}:

In the Appendix, we prove that any one-particle probability density for an ensemble of particles obeys the equation:
\begin{equation}
\frac{\partial \rho}{\partial t}+\nabla\cdot(\rho \textbf{v})=0\label{continuity0}
\end{equation}
here $\rho(\textbf{r},t)$ is the probability density for a particle, while $\textbf{v}(\textbf{r},t)$ is the velocity of a particle from the ensemble at a position $\textbf{r}$ and time $t$.

If we substitute $\textbf{v}$ from the guiding condition, we have that the probability density $\rho$ obeys the equation:
\begin{equation}
\frac{\partial \rho}{\partial t}+\nabla\cdot\left(\rho \frac{\hbar}{2mi}\frac{\Psi^{*}\nabla\Psi-\Psi\nabla\Psi^{*}}{\Psi^{*}\Psi}\right)=0\label{dens1}
\end{equation}

On the other hand we have proved that (see Eqs. \eqref{Schr123} and \eqref{Schr124}):
\begin{equation}
\frac{\partial}{\partial t}|\Psi|^{2}+\nabla\cdot\left[\frac{\hbar}{2mi}\left(\Psi^{*}\nabla\Psi-\Psi\nabla\Psi^{*}\right)\right]=0\label{dens2}
\end{equation}
These equations are always true. Now, at the initial moment $t=t_{0}$ Eq. \eqref{dens1} is:
\begin{align}
&\frac{\partial \rho}{\partial t}\Bigg|_{t=t_{0}}=\notag\\
&-\nabla\cdot\left(\rho(\textbf{r},t_{0}) \frac{\hbar}{2mi}\frac{\Psi^{*}(\textbf{r},t_{0})\nabla\Psi(\textbf{r},t_{0})-\Psi(\textbf{r},t_{0})\nabla\Psi^{*}(\textbf{r},t_{0})}{\Psi^{*}(\textbf{r},t_{0})\Psi(\textbf{r},t_{0})}\right)
\end{align}
But we can cancel the denominator with $\rho(\textbf{r},t_{0})$, since $\rho(\textbf{r},t_{0})=|\Psi(\textbf{r},t_{0})|^{2}$ and we have:
\begin{align}
\frac{\partial \rho}{\partial t}\Bigg|_{t=t_{0}}=-\nabla\cdot\left[\frac{\hbar}{2mi}\left(\Psi^{*}\nabla\Psi-\Psi\nabla\Psi^{*}\right)\right]\Bigg|_{t=t_{0}}=\frac{\partial}{\partial t}|\Psi|^{2}\Bigg|_{t=t_{0}}
\end{align}
At the last line we have used Eq. \eqref{dens2}. In other words, we reach the conclusion that:
\begin{equation}
\rho(\textbf{r},t_{0})=|\Psi(\textbf{r},t_{0})|^{2}
\end{equation}
as well as
\begin{equation}
\frac{\partial \rho}{\partial t}\Bigg|_{t=t_{0}}=\frac{\partial}{\partial t}|\Psi|^{2}\Bigg|_{t=t_{0}}
\end{equation}

These two equations tell us that in the next moment of time $t_{1}=t_{0}+\delta t$:
\begin{align}
&\rho(\textbf{r},t_{1})\approx\rho(\textbf{r},t_{0})+\delta t\frac{\partial \rho}{\partial t}\Bigg|_{t=t_{0}}\notag\\
&=|\Psi(\textbf{r},t_{0})|^{2}+\delta t\frac{\partial}{\partial t}|\Psi|^{2}\Bigg|_{t=t_{0}}\approx|\Psi(\textbf{r},t_{1})|^{2}
\end{align}
This approximate equation becomes true in the limit $\delta t\rightarrow 0$. This means, that the probability density $\rho(\textbf{r},t_{1})$ is the same as $|\Psi(\textbf{r},t_{1})|^{2}$ at the \textit{next} moment $t_{1}=t_{0}+\delta t$. But using the same logic, $\rho=|\Psi|^{2}$ even after $t_{1}$, at the moment $t_{2}=t_{1}+\delta t$ and also for $t_{3}=t_{2}+\delta t$, and so on. Therefore we reach the conclusion that the probability density $\rho=|\Psi|^{2}$ for \textit{all} moments of time $t>t_{0}$. This completes the proof of the theorem.

Pilot wave theory simply starts with the assumption that perhaps at the Big Bang, the particle had initial probability distribution $\rho(\textbf{r},t_{0})=|\Psi(\textbf{r},t_{0})|^{2}$. Then it is forever true that $\rho=|\Psi|^{2}$ for $t>t_{0}$.

In that case, the pilot wave theory explains the electron gun experiments in essentially the same way as the Copenhagen interpretation. However, instead of probability for a sudden collapse, we have probability for the particle to be guided at some location. At the interference minimums in Experiment 3 for example, there is $0$ probability for the particle to be guided there, since there $\Psi=0$. In the same way we explain the diffraction in Experiment 2. The wave function is not $0$ at the location of the widened and diffracted spot and there is non-zero probability for the particle to be guided there.

\section{Conclusion}
In this paper we have derived the famous Schr\"{o}dinger's equation from the classical Hamilton-Jacobi equation. We have followed several steps. First, we have showed that a light wave can behave as a beam of particles, in the so called regime of geometrical optics. In this regime, a beam of light does not penetrate into the geometrical shadow (no diffraction). In addition, in this regime the wave equation is reduced to the so called Eikonal equation. Second, we have derived the Hamilton-Jacobi equation in classical mechanics. Third, we have compared the Eikonal equation of geometrical optics and Hamilton-Jacobi equation in classical mechanics. We have used a specific approximation, which naturally leads from the Eikonal equation to the wave equation. Then we have applied this approximation, and thus we have derived Schr\"{o}dinger's equation from the Hamilton-Jacobi equation. Fourth, we have explained three particular electron gun experiments and have compared pilot wave theory of de Broglie (and later perfected by Bohm) with the Copenhagen interpretation. To our knowledge, this derivation of Schr\"{o}dinger's equation as well as the particular approximation that leads from geometrical optics regime to the wave optics regime, has not been done before. We hope this derivation will lead to better understanding of quantum mechanics.

\section{Appendix}
In this Appendix we prove the continuity Eq. \eqref{continuity0} for the one-particle probability density $\rho(\textbf{r},t)$. To this end, let us choose an ensemble of particles with $N$ particles, where $N$ is a really large number, say Avogadro number. Then the number density of the particles will be given by:
\begin{equation}
\mathcal{N}(\textbf{r},t)= N\rho(\textbf{r},t)
\end{equation}
Let us pick some small volume $\Delta V=\Delta x\Delta y\Delta z$ centered at a point $\textbf{r}=(x,y,z)$, as shown in FIG. 17. At the end of the calculation we shall take the limit $\Delta V\rightarrow 0$.

\begin{figure}[tb]
\includegraphics[width= 1.0\columnwidth]{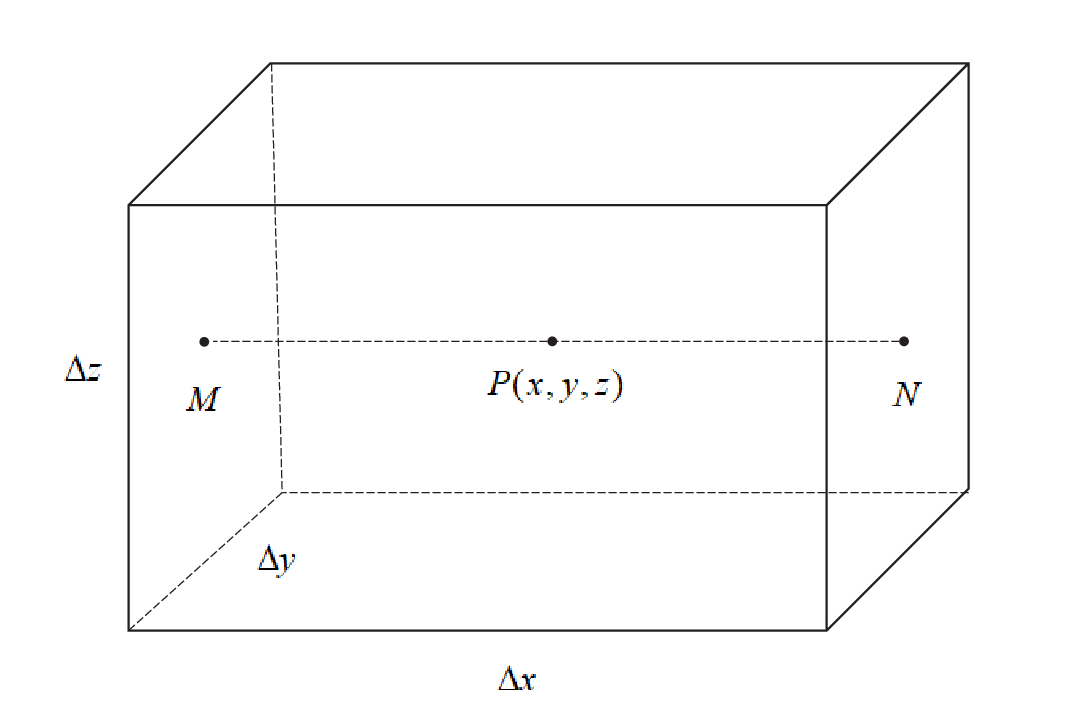}
\caption{}
\label{fig17}
\end{figure}

The number of particles that leave through the right side per unit time of the prism on FIG. 17 (point $N$) is $\mathcal{N}(x+\Delta x/2,y,z)v_{x}(x+\Delta x/2,y,z)\Delta y\Delta z$. If this number is \textit{negative}, the particles are \textit{entering} (not leaving) into the prism from the right side. The number of particles that leave through the left side of the prism per unit time on FIG. 17  (point $M$) is equal to $-\mathcal{N}(x-\Delta x/2,y,z)v_{x}(x-\Delta x/2,y,z)\Delta y\Delta z$. The total number of particles leaving the prism per unit time through the left and right side of the prism is then:
\begin{align}
\frac{\Delta N_{x}}{\Delta t}=&\mathcal{N}\left( x+\frac{\Delta x}{2},y,z \right) v_{x}\left( x+\frac{\Delta x}{2},y,z \right)\Delta y\Delta z\notag\\
&-\mathcal{N}\left( x-\frac{\Delta x}{2},y,z \right) v_{x}\left( x-\frac{\Delta x}{2},y,z \right)\Delta y\Delta z \notag\\
&\approx\frac{\left(\partial\mathcal{N}v_{x}\right)}{\partial x}\Delta x\Delta y\Delta z
\end{align}
In the last line we divided and multiplied by $\Delta x$ and used the fact that for some arbitrary function $g(x,y,z)$ we have that:
\begin{align}
& g\left( x+\frac{\Delta x}{2},y,z \right)-g\left( x-\frac{\Delta x}{2},y,z \right)\notag\\
&=\frac{g\left( x+\frac{\Delta x}{2},y,z \right)-g\left( x-\frac{\Delta x}{2},y,z \right)}{\Delta x}\Delta x\approx\frac{\partial g}{\partial x}\Delta x
\end{align}
We also add up $\Delta N_{y}/\Delta t$ and $\Delta N_{z}/\Delta t$ for the particles leaving the other sides and then the total number of particles, leaving the prism per unit time is:
\begin{align}
\frac{\Delta N}{\Delta t} &\approx\left(\frac{\left(\partial\mathcal{N}v_{x}\right)}{\partial x}+\frac{\left(\partial\mathcal{N}v_{y}\right)}{\partial x}+\frac{\left(\partial\mathcal{N}v_{z}\right)}{\partial x}\right)\Delta x\Delta y\Delta z \notag\\
&=\nabla\cdot\left(\mathcal{N}\textbf{v}\right)\Delta V
\end{align}
Here we have used that $\Delta V=\Delta x\Delta y\Delta z$. Taking the limit $\Delta V\rightarrow 0$ as well as $\Delta t\rightarrow 0$, we obtain the number of particles leaving the volume $dV$
\begin{equation}
\frac{dN}{dt}=\nabla\cdot\left(\mathcal{N}\textbf{v}\right) dV\label{1}
\end{equation}
On the other hand, the particles that left the volume $dV$ lead to decrease of its density:
\begin{equation}
\frac{dN}{dt}=-\frac{\partial \mathcal{N}}{\partial t}dV\label{2}
\end{equation}
From Eqs. \eqref{1} and \eqref{2} we derive:
\begin{equation}
-\frac{\partial \mathcal{N}}{\partial t}=\nabla\cdot\left(\mathcal{N}\textbf{v}\right)
\end{equation}
If we divide this equation by $N$ and if we take into account that $\mathcal{N}/N=\rho$, we finally derive the continuity equation:
\begin{equation}
\frac{\partial \rho}{\partial t}+\nabla\cdot\left(\rho \textbf{v}\right)=0
\end{equation}

\section{Acknowledgment}
This research did not receive any specific grant from funding agencies in the public, commercial, or not-for-profit sectors.

\section{Data Availability Statement}
The author confirm that the data supporting the findings of this study are available within the article and/or its supplementary materials.


\end{document}